\documentclass[useAMS,usenatbib]{mn2e}
\usepackage{graphicx,fleqn,multirow}
\DeclareGraphicsExtensions{.eps,.ps,.eps.gz,.ps.gz,.eps.Z}
\def \apj{ApJ}
\def \apjl{ApJL}

\def\gtsim{\raisebox{-.5ex}{$\;\stackrel{>}{\sim}\;$}}

\newcommand{\sw}{{\it Swift}}

\input psfig.sty

\title[
GRB100814A: an interplay of forward and reverse shocks?]{The optical rebrightening of GRB100814A: an interplay of forward and reverse shocks?}
\author[]{Massimiliano De Pasquale$^{1,2,3}$, N. P. M. Kuin$^1$, S. Oates$^1$,  S. Schulze$^{4,5,6}$, Z. Cano$^{4,7}$, 
\newauthor C. Guidorzi$^8$, A. Beardmore$^9$, P. A. Evans$^9$,  Z. L. Uhm$^2$, B. Zhang$^{2}$,  
\newauthor  M. Page$^{1}$, S. Kobayashi$^9$,  A. Castro-Tirado$^{10}$, J. Gorosabel$^{10,11,12}$, T. Sakamoto$^{13}$,   
\newauthor T. Fatkhullin$^{14}$, S. B. Pandey$^{15}$, M. Im$^{16}$,  P. Chandra$^{17}$, D. Frail$^{18}$, H. Gao$^{2}$,  
\newauthor  D. Kopa\v{c}$^{7}$,  Y. Jeon$^{16}$,  C. Akerlof$^{20}$, K. Y. Huang$^{21}$, S. Pak$^{22}$, W.-K. Park$^{16,23}$, 
\newauthor  A. Gomboc$^{19,24}$,  A. Melandri$^{25}$, S. Zane$^1$, C. G. Mundell$^7$, C. J. Saxton$^{1,26}$,   
\newauthor  S. T. Holland$^{27}$, F. Virgili$^7$, Y. Urata$^{28}$, I. Steele$^7$, D. Bersier$^7$, N. Tanvir$^9$,
\newauthor V. V. Sokolov$^{14}$, A. S. Moskvitin$^{14}$\\
$^1$Mullard Space Science Laboratory, University College London Dorking, Holmbury St. Mary, Dorking Surrey, RH5 6NT, United Kingdom\\
$^2$Department of Physics, University~of~Nevada,~Las~Vegas, United States\\
$^3$INAF/IASF, Via Ugo La Malfa 153, 90146, Palermo, Italy\\
$^4$Centre for Astrophysics and Cosmology, Science Institute, University of Iceland, Dunhagi 5, 107 Reykjav\'ik, Iceland\\
$^5$Pontificia Universidad Cat\'{o}lica de Chile, Departamento de Astronom\'{\i}a y Astrof\'{\i}sica, Casilla 306, Santiago 22, Chile\\
$^6$Millennium Institute of Astrophysics, Vicu\~{n}a Mackenna 4860, 7820436 Macul, Santiago, Chile\\
$^7$Astrophysics Research Institute, Liverpool John Moores University, 146 Brownlow Hill, Liverpool, L3 5RF United Kingdom\\
$^8$Department of Physics and Earth Sciences, University of Ferrara, via Saragat 1, I-44122 Ferrara, Italy\\
$^9$University of Leicester, University Rd, Leicester LE1 7RH, United Kingdom\\
$^{10}$Instituto de Astrof\'{i}sica de Andaluc\'{i}a (CSIC), P.O. Box, E-18080 Granada, Spain \\
$^{11}$Ikerbasque, Basque Foundation for Science, Alameda de Urquijo 36-5, E-48008 Bilbao, Spain\\
$^{12}$Unidad Asociada Grupo Ciencia Planetarias UPV/EHU-IAA/CSIC, Departamento de F\'{\i}sica Aplicada I, E.T.S. Ingenier\'{\i}a, \\
Universidad \ del Pa\'{\i}s-Vasco UPV/EHU, Alameda de Urquijo s/n, E-48013 Bilbao, Spain\\
$^{13}$NASA Goddard Space Flight Center, Greenbelt, MD 20771, USA\\
$^{14}$Special Astrophysical Observatory, Russian Academy of Science, Russia\\
$^{15}$ARIES, Manora Peak, Nainital, Uttarakhand, India, 263129\\
$^{16}$CEOU-Astronomy Program, Dept. of Physics \& Astronomy, FPRD, Seoul National University, Seoul, Republic of Korea\\
$^{17}$National Centre for Radio Astrophysics, Tata Institute of Fundamental Research, Pune University, Ganeshkhind, Pune 411 007, India\\
$^{18}$NRAO, P.O. Box 0, Socorro, NM 87801, USA\\
$^{19}$Faculty of Mathematics and Physics, University of Ljubljana, Jadranska 19, SI-1000 Ljubljana, Slovenia\\
$^{20}$500 East University, Ann Arbor, University of Michigan, MI 48109-1120 USA\\
$^{21}$Department of Mathematics and Science, National Taiwan Normal University, Lin-kou District, New Taipei City 24449, Taiwan\\
$^{22}$Kyung Hee University, 1 Seocheon-dong, Giheung-gu, Yongin-si Gyeonggi-do 446-701, Republic of Korea\\
$^{23}$Korea Astronomy \& Space Science Institute, 776 Daedukdae-ro, Yuseong-Gu, Daejeon 305-348, Republic of Korea\\
$^{24}$Centre of Excellence Space-SI, A\v{s}k\v{c}ereva cesta 12, SI-1000 Ljubljana, Slovenia\\
$^{25}$INAF - Brera Astronomical Observatory, via E. Bianchi 46, I-23807, Merate (LC), Italy\\
$^{26}$Physics Department, Technion - Israel Institute of Technology, Haifa 32000, Israel\\
$^{27}$Space Telescope Science Institute, Baltimore, 3700 San Martin Dr. Baltimore, MD 21218, USA\\
$^{28}$Institute of Astronomy, National Central University, Chung-Li 32054, Taiwan\\
}

\begin{document}

\date{Accepted...Received...}

\maketitle

\label{firstpage}

\begin{abstract}
We present a wide dataset of $\gamma$-ray, X-ray, UVOIR, and radio observations of the \sw~GRB100814A. At the end of the slow decline phase of the X-ray and optical afterglow, this burst shows a sudden and prominent rebrightening in the optical band only, followed by a fast decay in both bands. The optical rebrightening also shows chromatic evolution. Such a puzzling behaviour cannot be explained by a single component model. We discuss other possible interpretations, and we find that a model that incorporates a long-lived reverse shock and forward shock fits the temporal and spectral properties of GRB100814 the best. 
\end{abstract}


\section{INTRODUCTION} \label{intro}
Research on gamma-ray bursts (GRBs) has greatly benefitted of the \sw~mission (\citealt{ge04}). This space observatory carries three scientific instruments: the Burst Alert Telescope (BAT; Barthelmy et al. 2005), the X-ray telescope (XRT; \citealt{ba05}), and the Ultra-Violet Optical Telescope (UVOT; \citealt{rom05}). When BAT detects a GRB, {\it Swift} slews towards the source position within 1-2 minutes, and follows up the GRB afterglow emission  \citep{cos97, gal98} until it becomes too weak to be detected, usually a few days after the trigger. {\it Swift} also delivers the position of a newly discovered source promptly to ground based observatories, which can observe the optical and radio afterglows in bands and sensitivities which cannot be achieved by the space facility. Therefore, GRB observations in the {\it Swift} age cover the temporal behaviour of GRBs in many different electromagnetic bands from $\sim100$~s after the trigger onwards. 
Moreover, \sw~has dramatically increased the statistics of GRB afterglows observed (about 90 GRBs per year) from the past. Such comprehensive coverage and statistics have shown that the light curves of GRBs at different wavelengths can be surprisingly diverse. During the afterglow, changes of the flux decay-rate or even rebrightenings can occur in some electromagnetic bands but not in others. An obvious example is the X-ray flares, which do not usually show an optical counterpart (Falcone et al. 2006). Conversely, a few authors have examined GRBs with episodes of optical rebrightening which have no clear equivalent in the X-ray band, such as GRB081029 (Nardini et al. 2011, Holland et al. 2012), and GRB100621A (Greiner et al. 2013). Another less clear-cut case may be GRB050401 (De Pasquale et al. 2006). These events are particularly puzzling since, after the optical rebrightening, the X-ray and optical light curves resume similar behaviour, with simultaneous change of slope. 
This has called for a deep revision of the emission models of GRB afterglows, which in the past mostly involved a single emission component. Observations indicate that a single component cannot be responsible for the observed features, but all the components producing the afterglow may still be connected, and possibly have a common origin.
According to the most accepted scenario, the initial phase of $\gamma$-ray emission arises when dissipation process(es) occur in ultra-relativistic shells emitted by a central engine (Rees \& M\'esz\'aros 1994). The afterglow arises when the burst ejecta interact with the surrounding medium and produce two shocks; one moving forward in the medium (forward shock, or FS) and another one inward into the ejecta (reverse shock, or RS), causing their deceleration (M\'esz\'aros \& Rees 1993; Sari \& Piran 1999). Both shocks energize the electrons of the medium in which they propagate. The electrons in turn cool by synchrotron emission and produce the observed afterglow light. 
It is therefore possible that FS and RS can jointly contribute to the observed emission and, since their emissions peak at different wavelengths, produce the puzzling chromatic behaviour observed (e.g. Perley et al. 2014, Urata et al. 2014). Other scenarios put forward involve a residual `prompt' emission producing the X-rays (Ghisellini et al. 2007), up-scattering of the photons produced by FS by fast ejecta (Panaitescu 2008), evolution of the physical parameters of the blast waves (Panaitescu et al. 2006), and two-component jet (De Pasquale et al. 2009; Liang et al. 2013).

In this article, we present an ample dataset of the {\it Swift} GRB100814A and discuss the remarkable temporal properties of this event. GRB100814A shows a conspicuous rebrightening in the optical bands between $\sim15$ and $\sim200$~ks after the burst trigger. Such a rise of the optical flux has no clear counterpart in the X-ray light curve. However, the flux in both bands shows a similar quick decay after 200 ks. Radio observations show a broad peak about $10^6$~s after the trigger, followed by a slow decay which is different from the rapid fall of the flux visible in the X-ray and optical at the same epoch. Finally, we mention other {\it Swift} GRBs that show comparable features and how the modeling adopted in this paper might be applied to their cases.

 Throughout this paper, we use the convention $F_{\nu} \sim t^{-\alpha} \nu^{-\beta}$, where $F_{\nu}$ is the flux density, $t$ is the time since the BAT trigger, $\nu$ the frequency, $\alpha$ and $\beta$ are the temporal and spectral indices. The errors indicated are at $1~\sigma$ confidence level (68\%~C.L.), unless otherwise indicated.

\section{Reduction and analysis of data}

\subsection{{\it Swift} $\gamma$-ray data}

\begin{figure*}
 \centering
 \includegraphics[width=0.75\textwidth, angle=-00]{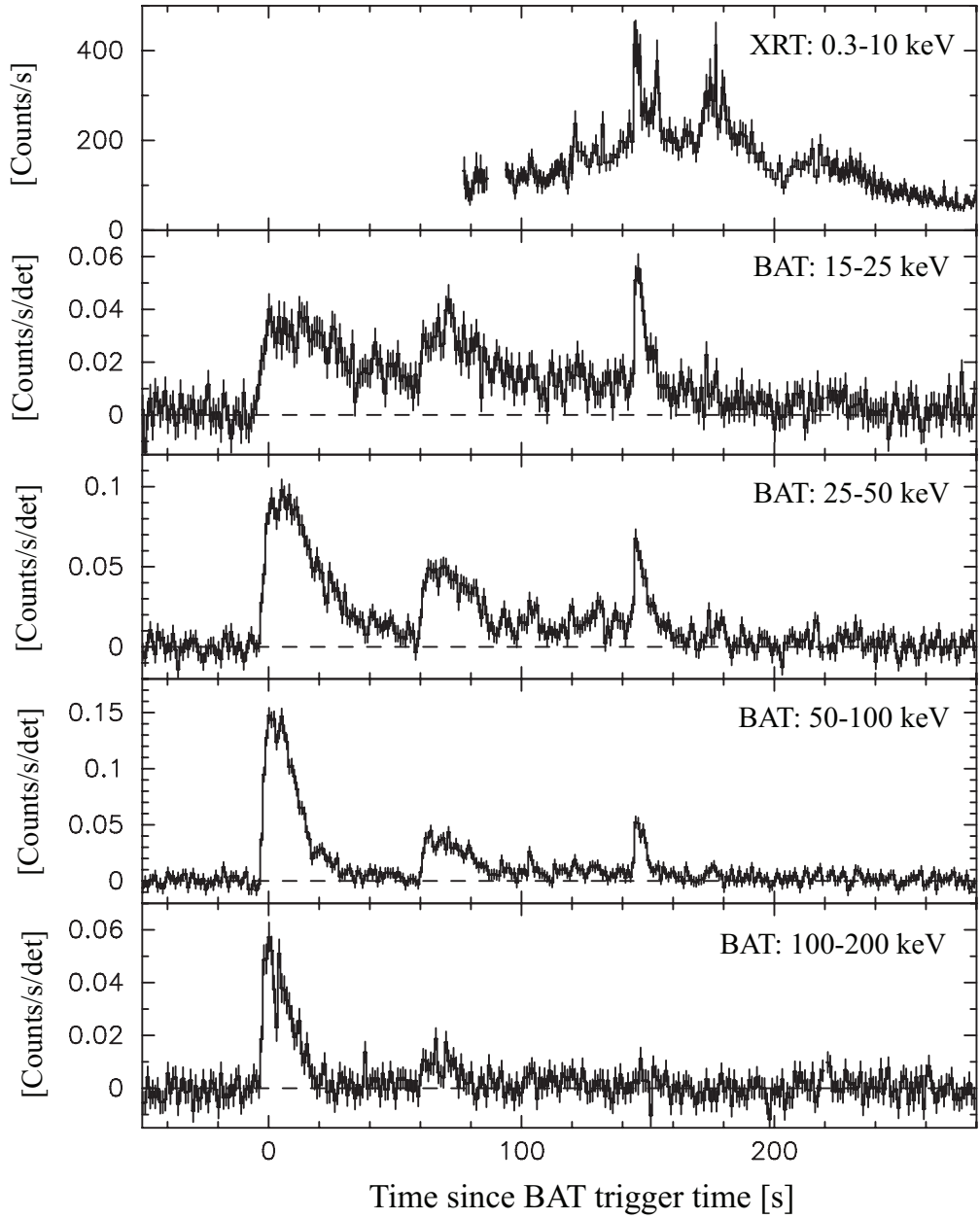}
 \caption{GRB100814A prompt emission detected by BAT and XRT.}
\label{100814a_prompt_lc}
\end{figure*}

 GRB100814A triggered the BAT instrument at T$_0$ = 03:50:11 UT on August 14, 2010 (\citealt{bea10}). The refined BAT position is R.A. (J2000) =$01^{\rm h} 29^{\rm m} 55^{s}$, Dec. (J2000) =-$17^{\circ} 59' 25.7''$ with a position uncertainty of 1' (90\% C.L., Krimm et al.  2010). The GRB onset occurred 4 seconds before the BAT trigger time and it shows 3 main peaks (see Fig.~\ref{100814a_prompt_lc}).

 From the ground analysis of the BAT data ($15-350~\mathrm{keV}$ energy band) we found that the GRB duration is {\it T}$_{90}=174.5 \pm 9.5$~s by {\tt battblocks} (v1.18). As for the spectral analysis, we will only consider results obtained in the $15-150~\mathrm{keV}$ band, because the mask weighted technique was used to subtract the background. In this case, it is not possible to use the data above $150$~keV where the mask starts to become transparent to the radiation. The BAT spectrum was extracted using {\tt batbinevt} (v1.48). The time-averaged spectrum from T$_0$-3 to T$_0$+235 s is best fitted by a simple power law model. The photon index is $1.47 \pm 0.04$ (90\% C.L.). This value is between the typical low energy photon index, $\simeq1$, and the high energy photon index, $\simeq2$ of GRB prompt emission described by the Band model \citep{ban93}. This suggests that that peak energy $E_{\mathrm{peak}}$ is likely to be inside the BAT energy range. The time-averaged $E_\mathrm{peak}$ is estimated to be $110_{-40}^{+335}~\mathrm{keV}$ using the BAT $E_\mathrm{peak}$ estimator (Sakamoto et al. 2009)  The fluence in the $15-150~\mathrm{keV}$ band is $(9.0 \pm 1.2) \times 10^{-6}$ erg cm$^{-2}$. The BAT 1-s peak photon flux is $2.5 \pm 0.2$ ph cm $^{-2}$ s$^{-1}$ in the $15-150~\mathrm{keV}$ band.  This corresponds to a peak energy flux of $(2.8 \pm 0.2) \times 10^{-7}$ erg cm$^{-2}$ s$^{-1}$ ($15-150~\mathrm{keV}$).  This 1-s peak  flux is measured from $T_0$(BAT) $-$0.06 s.  {\it Swift} began to slew to repoint the sources with the XRT and UVOT 18 seconds after the trigger, when the prompt emission had not yet ended.

 The prompt emission of GRB100814A was also detected by {\it Konus-Wind} \citep{gol10}, {\it Fermi} \citep{vok10}, and {\it Suzaku/WAM} \citep{nis10}.

 As observed by {\it Konus}, the event had a duration of $\sim150$ seconds and fluence of $(1.2\pm0.2) \times 10^{-5}$ erg cm$^{-2}$ in the 0.02-2 MeV band (90\%~C.L.).  The spectrum is best fitted by a power law plus exponential cut off model. The best fit parameters are a low-energy photon index $\Gamma_{1}=0.4 \pm 0.2$, and a cut off energy $E_{\mathrm{p}}=128\pm 12~\mathrm{keV}$. The value of this parameter is similar to $E_{\mathrm{peak}}$ drawn from BAT data. Assuming a redshift of $z=1.44$ \citep{ome10} and an isotropic emission, this corresponds to a $\gamma$-ray energy release of $\simeq 7\times10^{52}$ erg between 1 and 10000 $\mathrm{keV}$ in the cosmological rest frame of the burst. We derived this value using the $k$-correction of Bloom et al. (2001).

\subsection{X-ray data}
XRT initially found an uncatalogued bright X-ray source $48"$ from the BAT position. The ground-processed coordinates are R.A. (J2000) = $01^{\rm h} 29^{\rm m} 53.54^{\rm s}$, Dec.(J2000) = -17$^{\circ}$ 59' 42.1'' with an uncertainty of 1\farcs5 (90 \% C.L.). This source subsequently faded, indicating that it was the X-ray counterpart of GRB100814A. Windowed Timing (WT) mode data (with ms time resolution but only 1-D spatial information) were gathered up to $600$~s after the trigger, after which the data were gathered in Photon Counting (PC) mode (with 2.5-s time resolution and 2-D spatial information). For both the spectral and temporal analysis, we considered counts within the $0.3-10~\mathrm{keV}$ band.

For the temporal analysis, we used the automated XRT GRB light curve analysis tools of Evans et al. (2009, 2007). At late times, we noticed  the presence of a nearby source $11"$ away from the GRB position,  contributing a count-rate of $\sim 8\times 10^{-4} \thinspace {\rm counts \thinspace s^{-1}}$ (corresponding to a $0.3-10\thinspace {\rm keV}$ flux of $\sim 4.5 \, \times 10^{-14} {\rm erg \thinspace cm^{-2} \thinspace s^{-1}}$), which caused the light curve to flatten to a roughly constant level beyond $\sim 9\times 10^5$~s  after the trigger. To mimimise the effect of this nearby source on the GRB light curve at late times (after $2\times 10^5 \thinspace {\rm s}$) we used a fixed position extraction region (to prevent the automatic analysis software centroiding on the non-GRB source location), with a reduced extraction radius (of 23 arcsec) and ignored the data beyond $9\times10^5$~s after the trigger. The count-rate light curve was converted to a flux density light curve at 10 keV following Evans et al. (2010), which accounts for spectral evolution as the burst decays.

 Fig. 2 shows the X-ray light curve of GRB100814A, as well as the UV/optical/IR and radio ones. At the beginning of the XRT light curve we clearly distinguish a sequence of flares, the last one peaking at $\sim 220$~s, followed by a steep decay with slope $\alpha=4.65\pm0.08$, which we interpret as the end of the prompt emission phase. Unfortunately observations made during the first orbit end at $\sim 750$~s, which limits our ability to better define this phase of the emission, although the last data points seem to show a flattening of the light curve.
During the second orbit observations, starting at $\sim 3000$~s, the flux decays at a much slower rate. This phase seems to last until $\sim 10^5$~s, when the decay of the X-ray flux becomes much steeper. This second phase of steep decay ends at $\sim 9\times10^5$~s after the trigger, followed by a phase of roughly constant flux. This flux, however, is not due to the GRB afterglow, but to the unrelated source 10 arcseconds from the burst position.

The presence of a break at late time is obvious: if we try to fit the $0.3-10~\mathrm{keV}$ light curve from the beginning of the second orbit to $9\times10^5$~s with a single power law we get an unsatisfactory result ($\chi^2_{\nu}=494.5/306$ degrees of freedom, d.o.f.), while the use of a broken power law ($A t ^{-\alpha1}$ for $t\leq t_{\rm b}$; $A t_{\rm b}^{\alpha_{2} - \alpha_{1}} t^{-\alpha2}$ for $t\geq t_{\rm b}$) gives a very significant improvement ($\chi^2 =183.8/304$ d.o.f.). In this case, the best fit parameters are: decay indices $\alpha_{1}=0.52\pm0.03$ and $\alpha_{2}=2.11^{+0.15} _{-0.13}$, $t_{\mathrm{break}} = 133.1^{+7.9} _{-6.4}$~ks. In both cases, we added a constant to the broken power law model to take into account the presence of the serendipitous source. We also tried to fit the light curve with a smoothly joined broken power law model \citep{beu99}, which enables us to examine different ``sharpness'' of the X-ray light curve break. We have found that the data do not  discriminate between a smooth and a sharp transition. If all parameters are allowed to vary, a model with a sharp break ($n=10$ in the Beuermann et al. 1999 formula) produces a marginally better fit. We also note that the possible ``dip" at $\sim 6\times10^4$~s is not statistically significant.

 We fitted the PC spectral data from the second orbit up to $9\times10^5$~s after the trigger with an absorbed power law model, by accounting separately for the Galactic and intrinsic absorption columns (the latter at $z=1.44$). The Galactic column density was fixed at $1.75\times 10^{20}$~cm$^{-2}$ \citep{kal05}. The fit is statistically satisfactory ($\chi^2 _{\nu} = 182.3/218$ d.o.f.). The best fit value for the column density of the extragalactic absorber is N$_{\mathrm{H}}= 1.18 ^{+0.29} _{-0.28} \times 10^{21}$~cm$^{-2}$, which is significantly different from zero, and the energy index of the power law is $\beta_\mathrm{X}=0.93\pm0.03$.  We find no evidence for spectral evolution: parameters consistent with those given above are obtained when fitting the spectra taken before and after the $133$~ks break. When compared to other X-ray afterglows detected by {\it Swift}, the X-ray afterglow of GRB100814A has an average flux around $10^4$~s. However, the long X-ray plateau makes GRB100814A move to the bright end of the flux distribution at $\sim0.5$ day after the trigger in the cosmological rest frame (Fig. \ref{ag_lc.eps}). The $0.3-10~\mathrm{keV}$ X-ray flux normalized at 11 hr after the burst is $\simeq 10^{-11}$ erg cm$^{-2}$ s$^{-1}$ in the observer's frame.

\subsection{UVOT and ground optical observatories data.}\label{opticaldata}

{\it Swift}/UVOT observations started 77~s after the trigger, with a
11~s exposure taken in the {\it v} band while the spacecraft was still
slewing. A grism exposure followed, from which we derive a {\it b} magnitude \citep{kuin2015}. The first settled imaging
exposure, in the {\it u} filter, started 153~s after the trigger and lasted
250~s; this exposure was obtained in event mode so that the position
and arrival time of each photon was recorded. Immediately afterwards,
UVOT took a sequence of 20~s exposures, cycling through its colour
filters. After the first orbit of {\em Swift} observations finished at
$\sim$ 700s UVOT switched to longer cadence observations, including the
{\it white} filter in its sequence.

GRB100814A was observed with 10 different ground-based optical
telescopes (see Table~1) in a range of photometric bands. To minimise
systematics between the different observatories and bands, where
possible the same stars in the field surrounding GRB100814A were used
as secondary standards for the different photometric bands and
instruments. There are some practical limitations to this approach:
the fields of view of some instruments are smaller than the basic set
of secondary standards and the sensitivities of the instruments are
limited to different brightness ranges. That means that usually a
subset of the calibration stars was used for a particular instrument,
and sometimes additional calibration stars were used to complement the
common set.  The secondary standards were calibrated in {\it B} and {\it V} using
the UVOT {\it b} and {\it v} observations and the transformation equations
provided by \citet{Poole}. The secondary standards were calibrated in
$R$, $r'$ and $i'$ using the CQUEAN observations (see below), using 
the transformations
from \citet{jordi06} to obtain $R$ magnitudes. The $r'$ and $i'$ magnitudes of
the secondary standards were verified using the 1-m Lulin Optical
telescope observations (see below). The photometric errors which were 
assigned to the data include  
both the random error and the systematic error from the calibration of 
the secondary standards.

The GRB was observed with the CQUEAN instrument \citep{park12,kim11}
mounted on the McDonald 2.1m Otto Struve telescope for five nights.
During that time observations of two SDSS photometric standards, BD+17
4708 and SA113-260 \citep{JSmith} were obtained, and used to calibrate
both the GRB photometry and the surrounding field stars down to the
22nd magnitude. 

The GRB was also followed with the 1-m Lulin Optical Telescope
(LOT; Kinoshita et al. 2005, Huang et al. 2005). The secondary standards in the LOT field were
calibrated independently of the CQUEAN observations, using LOT
observations of four SDSS fields at a range of airmass on
Sept 14, 2010. The magnitudes of the secondary standards were cross
checked with the corresponding CQUEAN magnitudes and were found to be
consistent within the errors. 

Observations with the Robotic Optical Transient Search
\citep[ROTSE;][]{Akerlof} IIIc site, located at the H.E.S.S. site at
Mt Gamsberg, Namibia, were obtained starting 290~s after the trigger
time.  Unfortunately, the light of a nearby variable star (position
R.A.(J2000)=$1^{\rm h} 29^{\rm m} 53.978^{\rm s}$, Dec.(J2000)=-17$^{\circ}$ 59' 35.5", USNO R2=19.58 mag)
contaminated the observations past 1000~s. We removed the
star which caused problems by using image subtraction, to confirm the data prior to 1000~s are
valid and uncontaminated. 
Although the observations were taken without an optical filter, the peak
response is in the R band, and the ROTSE data were calibrated against
the USNO-B1 {\it R2} magnitudes of 29 sources within $10'$ of the
transient. 

We obtained late-time GRB observations with the
Scorpio instrument \citep{scorpio} mounted on the Russian BTA 6-m
telescope which were calibrated using the CQUEAN secondary
standards.  

Observations from the Liverpool Telescope \citep[LT]{LT} and Faulkes
Telescope North (FTN) were calibrated using a subset of the CQUEAN
secondary standards which are within the LT and FTN fields of view. 

GRB100814A was also observed in the $R$ band with 
the 1.23-m Calar Alto Astronomical
Observatory (CAHA), the
IAC-80 Telescope of the Observatory del Teide, Tenerife,
 and the Gran Telescopio Canarias (GTC), La Palma. 
In all three cases the data were calibrated using
the CQUEAN secondary standards.  

The GRB was also observed in $R$, $V$, and $B$ bands with the Northern
Optical Telescope (NOT) in La Palma. The images were calibrated using
the CQUEAN and UVOT secondary standards.

\begin{figure*}
 \centering
 \vspace{-1cm}
 \includegraphics[width=0.775\textwidth, angle=90]{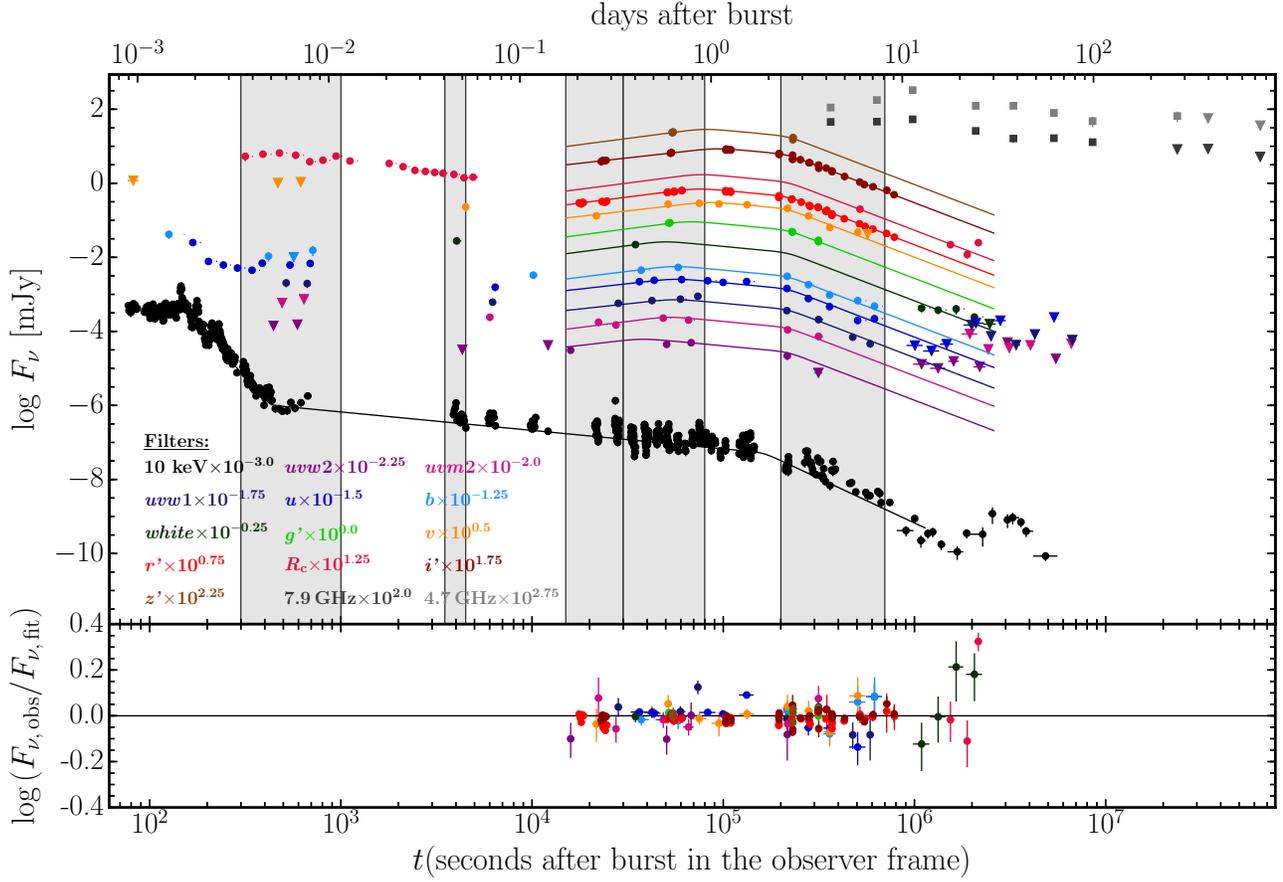}
 \caption{The afterglow GRB10814A from $10^{-5}~\mathrm{eV}$ to $1.73~\mathrm{keV}$. Individual light curves were shifted according to the text in the figure to separate the different light curves. Radio and optical upper limits are at $3\sigma$, and are shown as triangles. The X-ray data points after $10^6$ s are contaminated by an unrelated serendipitous source. The shaded areas point to the epochs when data for the SEDs were collected.}
\label{all_LCs}
\end{figure*}

\begin{figure*}
 \centering
 \includegraphics[width=0.75\textwidth, angle=-90]{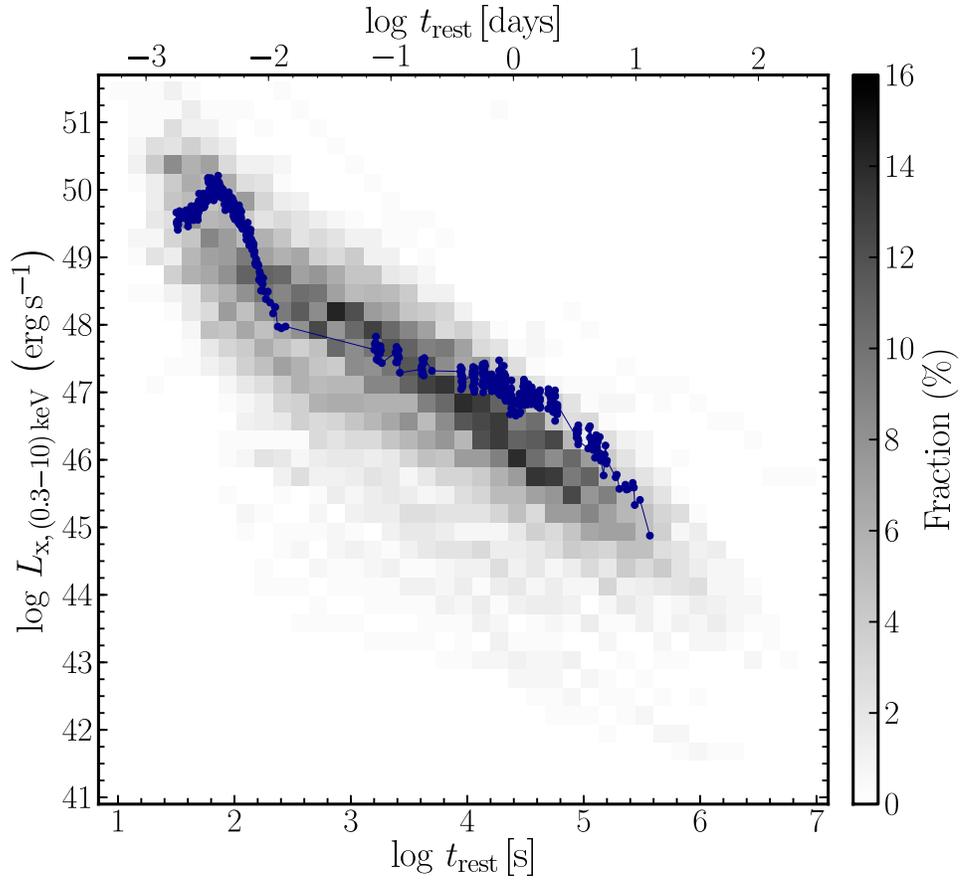}
 \caption{The afterglow luminosity of GRB 100814A between 0.3 and 10 $\mathrm{keV}$ in the cosmological rest frame is outlined in blue colour. It is also compared to
 the ensemble of {\it Swift} GRBs up to 2010, which is shown in shades of gray depending of the frequency of events having a certain flux at given epoch.}
\label{ag_lc.eps}
\end{figure*}

The resulting optical light curves are shown in
Fig~\ref{all_LCs}. In order to improve our understanding of the
behaviour of the optical light curve and check for the presence of
chromatic evolution of the emission, we followed two approaches. In
the first approach we normalised all of the light curves to a
single filter, and in the second approach we analysed the light curves
in different bands separately. The first approach, as described in \citet{oat09}, consists of renormalizing the light curves to a single filter.

\begin{figure*}
 \centering
 \vspace{0cm}
 \includegraphics[width=0.65\textwidth, angle=-90]{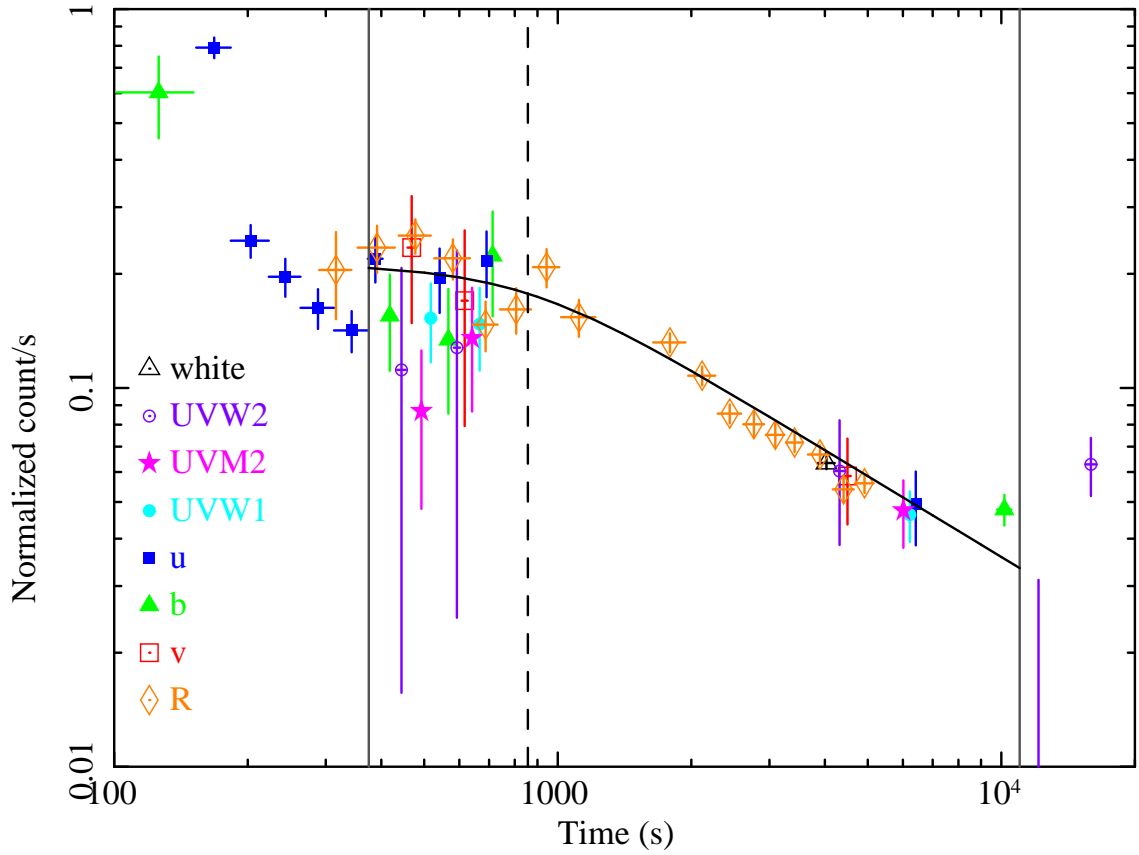}

\caption{Early optical light curves of GRB100814A. The vertical lines represent the time interval for the temporal fit of the early optical data (see section 2.3). The best fit model, a broken power law, is imposed over the data points. The dashed vertical line represents the break time of the broken power law fit (see Sect. 2.3). }



\label{fig:early_LCs}
\end{figure*}

The very early optical light curve varies rapidly. An optical flare peaks at $\sim 180$~s and then rapidly decays, basically giving no contribution after $\simeq 375$~s. After this early flare, we have a phase in which the optical flux is roughly constant, followed by a decay starting at $\sim 1000$~s in all filters. To investigate the early plateau and the following decay more throughly, we have renormalized the early data to the \sw~$u$ band filter. We then fitted the $375-11000$~s data points with a smooth broken power law, and the best fit parameters are $\alpha_{\mathrm{opt,2}}=0.03^{+0.16} _{-0.20}$, break time $t_{\mathrm{break}}=856^{+260} _{-190}$~s, $\alpha_{\mathrm{opt},3}=0.72\pm0.06$, with $\chi^2/\mathrm{d.o.f} = 52.5 / 33$. We show in Fig. \ref{fig:early_LCs} the renormalized early optical light curves.

The initial optical flare may be produced by the same process responsible for the early flaring activity in the X-ray, since the temporal behaviour is roughly similar. Flares are likely produced by internal dissipation mechanisms, such as internal shocks, which occur when the ultrarelavitistic ejecta shells interact with each other. It is possible that the plateau we see between $375$ and $1000$~s is due to a decreasing emission from internal dissipation and rising emission from the  external shock. Alternatively, the plateau might be due to a slow rise of the external shock emission only. We note that an initial plateau or shallow decay phase are associated to the external shock onset, as observed in several $\it{Swift}$ bursts \citep{oat09}. The origin of external shock emission is different from that due to internal dissipation mechanism. External shocks are produced by the interaction between ejecta and the circumburst medium and are likely to produce the long-lived and slowly varying afterglow emission. 
After $t_{\mathrm{break}}$, the optical flux follows the typical power law decay of GRB afterglows; the deceleration of the leading shell of the ejecta must have occurred at this time or earlier. In the following, we will assume that break time $t_{\mathrm{break}}$ marks the deceleration time, and will investigate the GRB afterglow from this time onwards. We will return to how the results of this article are affected if the actual deceleration is slightly earlier. Since the study of internal dissipation mechanisms is not the goal of this paper, we will not discuss the initial optical flare any further. 


 Unfortunately, the sampling of our optical light curve between 10 and 20~ks is not good enough to ascertain precisely when the optical emission stops decaying and begins to rise. What we can say is that the rebrightening approximately started about 15~ks and culminated about 100~ks after the trigger, although there is no strong variation in flux between 50 and 200~ks, during which the light curves seem to form a plateau.

 Between 15~ks and 200~ks, during the optical rebrightening, the light curves do not seem to align. This might indicate that a process of chromatic evolution is taking place during the afterglow of GRB100814A. In more detail, the rebrightening appears to be bluer than the other portions of the light curves. These results indicate that the optical spectrum during the rebrightening is different from that found before and after the rebrightening. This feature reinforces the idea that the rebrightening is due to different emission components.

It is also possible that there is a chromatic evolution {\it during} the rebrightening itself. In fact, if we renormalize the light curves during the rise between 15 and 60~ks, the light curves during the plateau do not match one another, with the data points of the redder filters being systematically above those of the bluer filter. Conversely, the light curves of the rise do not match one another if we renormalize them in the interval between 50 and 200~ks. However, the plateau phase, although being redder than the rise one, still shows a spectrum which is bluer than that of the following fast decay. In summary, it is possible that the rebrightening spectrum gets redder with time. This trend is found in other GRBs, such as GRB120404A (Guidorzi et al. 2014).

 We renormalized the late optical light curves to the $i'$ band, since we have a good coverage in this filter in late observations. This technique was applied to data points between $250$~ks, when the fast decay has clearly started, and $10^6$~s. A fit with a power law model yields an acceptable result: $\chi^2$/d.o.f. = 82.4/53. It provides a best fit decay slope of $\alpha = 2.00 \pm 0.07$. After $10^6$~s, the optical emission was very weak and difficult to constrain. At the time, contamination from the constant flux of the host galaxy may also be possible. This has been accounted for in the fit of the late decay by adding a constant in the model. We note that Nardini et al. (2014) found a late decay slope of $\alpha = 2.25 \pm 0.08$ between 200 ks and $10^6$ s, using GROND data (taken in $g'r'i'z'$ and $J$, $H$ and $K$ bands) and including the contribution of the host galaxy; such a value is consistent with our best fit value above. By means of observations of the Calar Alto 3.5-m telescope, 3 years after the event, we determined that the host galaxy of GRB100814 shows a magnitude $J=22.32 \pm0.32$ (Vega; error including calibration uncertainties). 
 
In order to obtain a clearer insight into the event, we studied the single light curves during the rebrightening (from 15~ks onwards) as well. Based on the densest sampled light curves, we find that the late-time evolution is characterised by two breaks. All UV and optical light curves are fit with a smoothly double broken power law \citep{Liang2008a, Schulze2011a}, using Simplex and Levenberg-Marquardt algorithms \citep{Press2002a}. The uncertainties in the data were used as weights. First, the parameters for each light curve were set to be identical, except for the normalisation constants. The quality of the fit is bad ($\chi^2 _{\rm red}$ = 5.66 for 148 d.o.f.); furthermore the residuals in the first two power law segments are not randomly distributed and show a trend with wavelength. The residuals around the second break and in the third power law segment are small and randomly distributed around the fit, implying that the evolution during the first power law segment is either chromatic or that the model used is just not good enough, and that the evolution after the second break is achromatic.

\begin{figure*}
 \centering
 \includegraphics[width=0.75\textwidth, angle=90]{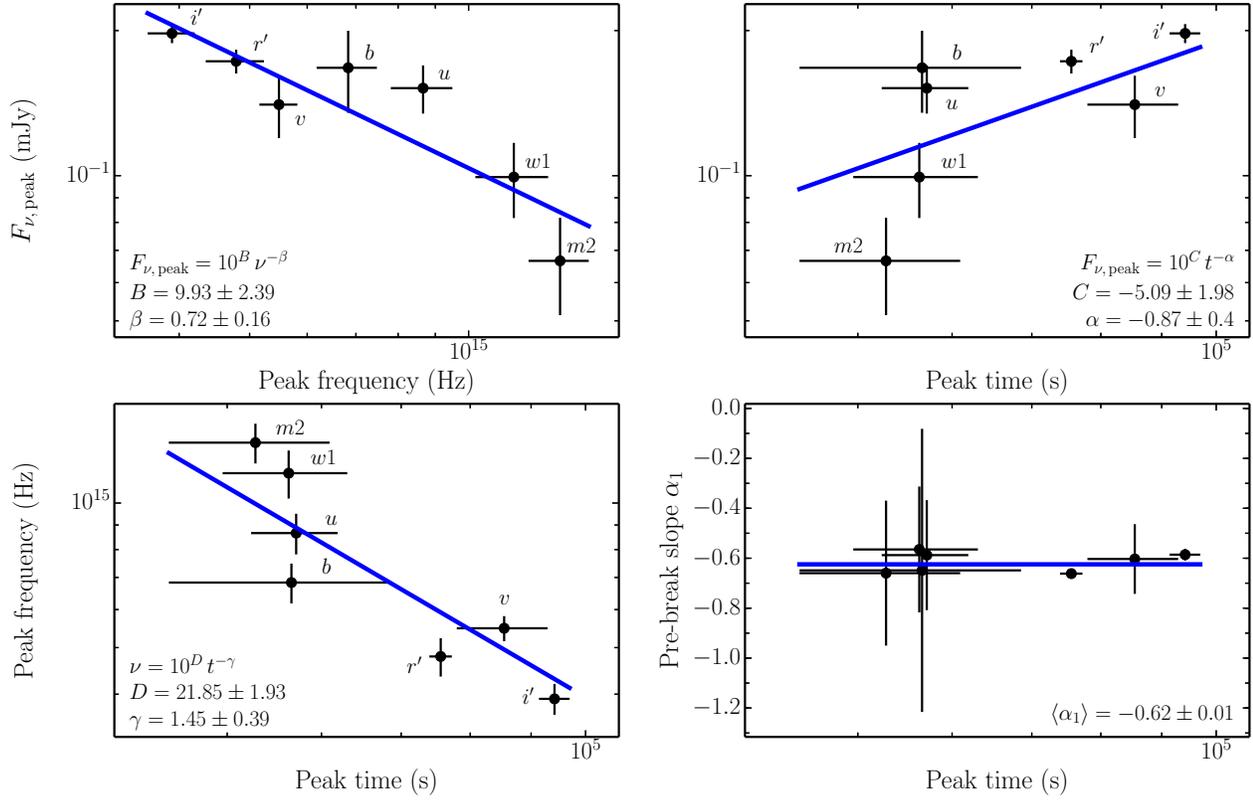}
 \caption{
 	Relations between the peak time, $t_{\rm peak}$, the peak flux density, $F_{\nu,\rm{peak}}$,
 	and the pre-peak slope, $\alpha_1$ in different UV and optical filters. The properties
 	of the fits are summarised in Table \ref{tab:ag_correlations}.
 	}
\label{fig:ag_correlation}
\end{figure*}

Next, we allowed the decay slope $\alpha_1$ and the break time $t_\mathrm{b,\,1}$ to vary for each band independently. Not only did the fit statistics significantly improve ($\chi^2 _{\rm red}$/d.o.f. = 2.96/126), but also the amplitude of the residuals decreased substantially. We summarise the fit parameters in Table \ref{tab:ag_fits}. The behaviour in the first power law segment is strongly frequency dependent, since the peak time\footnote{The peak time was computed from $dF_{\nu}/dt = 0$ (see \citet{Molinari2007a} for an explicit expression)} $t_\mathrm{peak}$, the peak flux density $F_{\nu,\,\rm p}$ are not the same for different frequencies $\nu$. To estimate the uncertainty, we only considered the error in the break time $t_{\rm{b},\,1}$.  The uncertainties of the first power law segment are for most of our data sets too large to detect any trend. Estimating the correlation and linear regression coefficients is not trivial, because the uncertainties in all parameters ($\alpha_1$, $t_{{\rm peak}}$, $F_{\nu,\,p}$, $\nu$) are not small. Owing to this, we applied a Monte Carlo technique \citep{Varian2005a}. In this method, every data point is represented by a 2D Gaussian, where the centre of peaks in each dimension are the parameter estimates, and the corresponding $1\sigma$ errors are the width of the distributions.
From these, we construct 10,000 resamples of the observed data sets, each of which is obtained by a random sampling with replacement from the original data set. For each of these data sets we compute the linear regression and correlation coefficients. The results are shown in Fig. \ref{fig:ag_correlation} and Table \ref{tab:ag_correlations}. 
From a statistical point of view, we do not find clear correlations. The most significant one is between $F_{\nu,\,\mathrm{p}}$ and $\nu$ with a correlation coefficient of -0.81 (Table \ref{tab:ag_correlations}), but even this correlation has significance of only $\simeq2\sigma$. The correlations between $F_{\nu,\,\mathrm{p}}$ and $t_{\rm peak}$, and $\nu$ and $t_{\rm peak}$ are not tight probably due to the large uncertainties in the break time. It is perhaps more correct to speak of `trends' rather than correlations but, thanks to some small error bars of the parameters we have derived, we can still safely state that the light curves in redder filters have higher peak fluxes and later peak times than those in the bluer filters. Any theoretical interpretation should explain this chromatic behaviour of the optical rebrightening.
 We note that the second break time, when the optical flux starts to decay fast, is not consistent with the X-ray late break time, although the decay slopes are consistent.
\\

\subsection{Radio data}\label{radio}
GRB100814A was observed with the Expanded Very Large Array (EVLA) in wide C-band receiver  with frequency at 4.5 and 7.9 GHz bands. The observations started on 2010 August 18 at 09:07 UT, $364.6$~ks  after the burst. Ten epochs were taken in total, with the last being 744 days after the trigger. The first 4 epochs of observations were in EVLA C configuration, whereas the fifth epoch of observations was in hybrid DnC configuration. The sixth and seventh epochs of observations were made in EVLA lowest resolution D configuration mode. The flux density scale was tied to the extragalactic source 3C48 (J0137+331), whereas J0132-169 was used as flux calibrator. The observations were made for 1 hour at each epoch, including the calibrators. The data were analysed using standard AIPS routines. The GRB was detected at all the first 6 epochs. At the seventh epoch on 2010 Nov 21 (about 8700 ks after the trigger), the radio afterglow was detected at 7.9 GHz, but it was not detected at 4.5 GHz. The afterglow was not detected in either band in the remaining epochs.
The peak flux was observed 11.32 days after the GRB. The peak flux densities were $582 \pm33$~$\mu$Jy and $534 \pm 27$~$\mu$Jy in the 4.5 and 7.9 GHz bands, respectively. The light curves in these two bands (visible in Fig. 2) show nearly simultaneous peaks, and their evolution afterwards looks similar, but the slopes before the peak different. We fitted both light curves with a smooth broken power law model, and we found the following best fit parameters: $\alpha_{4.5,1}= -1.27 ^{+0.20} _{-0.24}$, $t_{4.5,\mathrm{peak}} = 955.5^{+61.9} _{-56.0}$~ks, $\alpha_{4.5,2} = 0.89 ^{+0.11} _{-0.10}$; $\alpha_{7.9,1}= -0.19^{+0.12} _{-0.13}$, $t_{7.9, \mathrm{peak}} = 984.2^{+144.0} _{-116.2}$~ks, $\alpha_{7.9,2} = 0.77 ^{+0.09} _{-0.08}$. The two rise slopes are inconsistent at $\simeq5\sigma$, but the decay slopes are basically identical.

\section{Spectral Energy Distributions at several epochs}
\label{SEDs}
To constrain the spectral indices of the optical and X-ray emission, we built and fitted the spectral energy distributions (hereafter SEDs) of the X-ray and optical emission. We chose the epochs of 500~s, 4.5~ks, 22~ks, 50~ks, and 400~ks. The methods used to construct the SEDs are described in Schady et al. (2007). For the optical parts of the SEDs, the UVOT photometry has been supplemented with ground based photometry when available. For data taken in the $g'$, $r'$, $i'$ and $z'$ bands, response functions have been taken from \citet{fukugita96}. The $R$ band data which have been used in the SEDs come from the IAC 80 telescope, and so for these data the response function was based on the IAC 80 $R$ filter and CCD response\footnote{http://www.iac.es/telescopes/pages/en/home/telescopes/iac80.php}. We tried three fitting models, based on power law functions since the emission is synchrotron radiation. In the first one, the X-ray and optical were on the same power law segment. The second model is a broken power law. The third model is a broken power law with the difference between the spectral indices fixed to 0.5, as predicted in the case of a synchrotron emission cooling break.
In all fitting models, we added two components of absorption. The first component is due to our Galaxy and fixed at the value given by the Leiden/Argentine/Bonn Survey, $N_H = 1.8\times10^{20}\mathrm{cm}^{-2}$. The second component represents the extragalactic absorption, with the redshift fixed at $z=1.44$. Similarly, we added three components for the extinction. The first component represents the Galactic extinction, fixed at the value given by Schlegel et al. (1998), $E(B-V)  =0.02$~mag. The second component represents the extinction in the environment of the burst at redshift z=1.44; we chose the Small Magellanic Cloud extinction law, since it usually fits the extinction properties of the medium of GRB host galaxies (Schady et al. 2010). The third component is UV/optical attenuation by the intergalactic medium (Madau 1995).
 Since we do not detect any significant change in the X-ray spectrum from $\sim3000$~s to $\sim10^6$~s, we can assume that the X-ray spectral index is always the fairly constrained value determined using the whole dataset. Therefore, in all fits the spectral slope of the segment encompassing the X-ray band is forced between 0.84 and 1.02, i.e. within the best value of the fit of the X-ray data alone plus or minus 3$\sigma$. Given this constraint, no fits produced with a simple power law provide a statistically acceptable fit, with the exception of the 400~ks SED, and we do not consider them in the analysis below. The 500~s SED does not enable us to constrain fit results well, and we do not use it in our discussion. 

\begin{figure*}
 \centering
 \includegraphics[width=0.75\textwidth, angle=-90]{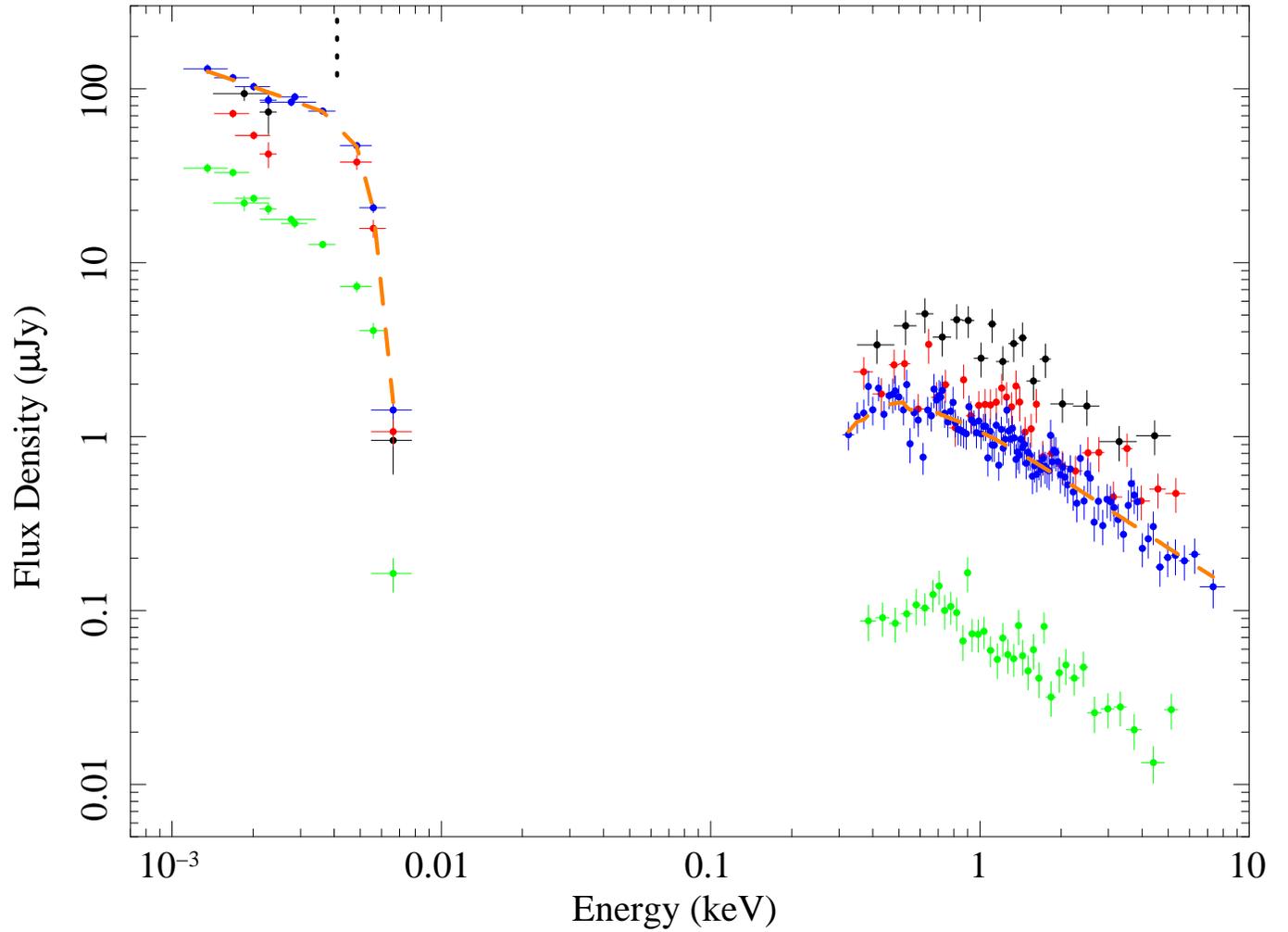}
 \caption{SEDs at 4.5 ks (black), 22 ks (red), 50 ks (blue) and 400 ks (green). The orange long-dashed line connecting the points of the 50 ks SED shows the best-fit model, that is the sum of two broken power laws; the black short-dashed vertical line indicates the position suggested for the synchrotron peak frequency of the first component. The 22 ks and 50 ks SEDs appear flat in the optical, while the 4.5 and especially the 400 ks SEDs show a steeper optical spectrum.
}
\label{fig:SEDs}
\end{figure*}

In the case of the 22 and 50~ks SEDs, we have also tried to fit the data with a model which is the sum of two broken power laws. This tested the possibility that two distinct components produce the optical and the X-ray flux and, given the chromatic behaviour of the optical afterglow, that the synchrotron peak frequency $\nu_{\mathrm{M}}$ is within or close to the optical band at these epochs. Thus, the low energy segment of the component producing the optical flux has been frozen to $\beta = -1/3$, while the component producing the X-ray flux has a break with differences in spectral slopes fixed to 0.5, as predicted by the external shock models (see Section 4). In the case of the $50$~ks SED, the sum of 2 broken power law models yields a slightly better fit than the model with a single broken power law and difference between the spectral indices fixed to 0.5: $\chi^2=111.6/112~\mathrm{d.o.f}$ versus $123.4 / 115~\mathrm{d.o.f}$. The best-fit break of the first component is $4.1^{+0.5} _{-0.6}$~eV. In the case of the $22$~ks SED, the fit becomes indistinguishable from a single broken power law model. We calculated the probability $P$ that the improvement in the fit of the 50~ks SED is given by chance by means of the F-test. We find that $P \simeq 1.1\times10^{-2}$. We tested these results by repeating the fit of the 50~ks SED with two broken power laws assuming Milky Way (MW) and Large Magellanic Cloud (LMC) extinction laws. In the case of the MW extinction law, the break of the first component is at $3.2^{+0.8} _{-0.4}$~eV, while the break is $4.8\pm1.0$~eV adopting a LMC extinction law. The two fits yield $\chi^2$ /d.o.f. = $106.9/112$ and $\chi^2$/d.o.f. = $107.9/112$. Fitting the 50~ks SED with a single broken power law model and difference between the spectral indices fixed to 0.5 with MW and LMC extinction laws yields $\chi^2$/d.o.f. = $113.1/115$ and $ \chi^2$/d.o.f. = 120.6/115, respectively.\\
Thus the results do not depend sensitively on the choice of extinction law. In conclusion, broken power law and two-broken power law models are perfectly acceptable for the 50~ks SED, but the model with two broken power law components is preferred by the data; one of the two breaks is found in or near the optical band. Results are summarised in Tab. \ref{tab_SEDs} and shown in Figure \ref{fig:SEDs}. The plot indicates changes in the spectral shape: while the 4.5 ks and the 400 ks SEDs show a normally steep optical spectrum, the 50 ks SED seems to have a flat optical emission. Furthermore, the 50 ks SED shows a steep optical-to-X index, which indicates that an additional optical component is needed with respect to other SEDs.

\section{Discussion}

The most remarkable property of GRB100814A is the broad optical peak which started roughly 15~ks after the trigger and ended at about 200~ks, followed by a steep decay with a rate similar to that observed in the X-ray band at the same time. The rebrightening is chromatic, since throughout it the X-ray light curve keeps decaying at the same rate as it did before and shows no obvious counterpart of the rebrightening. When fitting the SED built at the peak of the rebrightening, we find a break frequency in the optical band. We also find that the peak time and maximum flux evolve with the frequency. Later on the optical flux starts decaying faster, and roughly at the same time the X-ray flux began to decay with approximately the same temporal slope.

 This however leads to critical questions regarding the sources of the emission in GRB100814A: if the X-ray and the optical fluxes are due to the same component, why do they behave so differently with the optical showing a rebrightening? And if the optical rebrightening is due to a different component, why does it end at about the time of the steep break in the X-ray?\\

\subsection{Single component FS model}\label{single}

 In GRB100814A both the X-ray flux and optical light curves initially show a shallow decay. Slow early decay has been seen commonly in GRB afterglows (Liang et al. 2007), both in the X-ray and in the optical. Its origin is still a matter of debate. One of the most popular explanations is a phase of energy injection into the ejecta, which may be due to Poynting flux emitted by the burst central engine or trailing shells of outflow that collide with the leading parts of it (Zhang et al. 2006). The steep, late decay observed in both the X-ray and in the optical bands at the late epoch could only be attributed to a jet phase in the context of the FS model.

One can immediately check whether the standard FS model can explain the observed behaviour. The spectral and temporal indices of the flux of the observed bands are predicted by this model to be linked in relations which depend on the positions of the synchrotron self-absorption frequency $\nu_{\mathrm{SA}}$, the peak frequency $\nu_{\mathrm{M}}$ and cooling frequency $\nu_{\mathrm{C}}$ and the kind of expansion - collimated (jet) or spherical - and on the density profile of the surrounding medium, either constant (like in the interstellar medium, ISM) or decreasing with radius (like a stellar wind) (Sari, Piran \& Narayan 1998; Sari Piran \& Halpern 1999; Chevalier \& Li 2000; Kobayashi \& Zhang 2003a).

The only ways to account for the rise of the optical light curves are to assume a transit of $\nu_{\mathrm{M}}$ throughout the optical band, or the onset of the FS emission. The former would also explain the chromatic nature of the event. We note that we can fit the X-ray light curve as the sum of two components: one rapidly decaying, likely connected with the prompt emission, and a rising component that peaks at $\simeq 900$~s, and successively produces the slow decay observed. If we assumed that this time were the peak time and the X-ray frequency $\nu_{\mathrm{X}} = 4.2\times10^{17}$ Hz ($1.73~\mathrm{keV}$) were the peak frequency, we would find that even the X-ray is consistent with the extrapolation of the relation between these two quantities from the optical band (bottom-left panel of Fig. 5). The X-ray peak would be shifted at much earlier time due to its higher frequency, but the X-ray and the optical would obey the same trends and be produced by the same component. 

However, if $\nu_{\mathrm{M}}$ were approaching the optical band, one should observe a flux rise from the beginning of observations in the ISM case or a decrease as $t^{-1/4}$ decay slope for stellar wind (with a density profile of $r^{-2}$, where $r$ is the distance from the progenitor, Kobayashi \& Zhang 2003a). Neither of which are observed. Furthermore, to keep $\nu_\mathrm{M,\mathrm{FS}}$ in the optical band with a flat spectrum, one would require an extremely high value of kinetic energy of the ejecta (see Sect.s \ref{chromatic} and \ref{RS-FS1}).  The optical bump cannot even be the onset of FS emission in the context of single component scenario, because one should not see the observed decrease of the X-ray and optical flux before it.

The observed flux depends on parameters such as the fractions of blast wave energy given to radiating electrons and magnetic field $\epsilon_e$ and $\epsilon_B$, the circumburst medium density $n$, and the index of the power law energy distribution of radiating electrons $p$. A temporal evolution of such parameters might explain the observed behaviour. An example is a change of density of the environment $n$. For frequencies below the cooling break, the flux is proportional to $n$, while the flux in bands above the break does not depend on it. It is therefore possible that a rapid increase in $n$ causes an optical rebrightening and simultaneously leaves the X-ray flux decay unperturbed, as we observe. Does this explanation predict the spectral changes that we see in the GRB100814A rebrightening? Since $\nu_{\mathrm C} \sim n^{-1}$, one may think that $n$ could increase so much that $\nu_{\mathrm C}$ enters the optical band and changes the shape of the SED. However, several simulations have shown that the light curves do not show prominent rebrightening even if the blast-wave encounters an enhancement of density (Nakar \& Granot 2007, Gat et al. 2013).\\

We therefore conclude that a single component FS model cannot explain the GRB100814A observed behaviour. In the next section, we discuss a few multi-component models to interpret the behaviour of the afterglow of this burst.

\subsection{Two-component jet seen sideways}

In this model, the prompt emission, the early optical and X-ray afterglow emission is produced by a wide outflow, while the late optical rebrightening is due to emission from a narrow jet seen off-axis. The emission from the latter is initially beamed away from the observer, however as the Lorentz factor decreases, more and more flux enters the line of sight. Such a scenario has been already invoked \citep{gra05} to explain late optical rebrightening features, so in principle it could explain the behaviour of GRB100814A.
We note that Granot et al. (2005) interpret X-ray rich GRBs and X-ray flashes, which are events with peak energy of the prompt emission in the 10-100 and 1-10 keV ranges respectively, as GRBs seen off-axis. GRB100814A does not belong to such categories, having a peak energy above 100 keV. However, the shallow decay and the rebrightening feature of its afterglow may still be interpreted in the off-axis scenario.
We shall now determine in more detail whether this scenario is plausible.\\

\vspace{-0.5cm}
\subsubsection{Narrow jet}\label{Narrow}
A relativistic jet initially observed off-axis will naturally produce a rising light curve; the exact slope depends on the ratio between the off-axis angle and the opening angle. Looking at the synthetic light curves created by the code in ``afterglow library" of \citet{van10} we notice that a jet seen at $\theta_\mathrm{obs} \sim 3\theta_j$ produces a rise with slope $\alpha \simeq -0.65$, and an initial decay with slope $\alpha \simeq 0.45$, which are similar to those we observe at the optical rebrightening (see also Granot et al. 2005). In this context, the peak luminsity observed at $\theta_\mathrm{obs}$ is related to that on axis by the formula

\begin{equation}
L_{\theta_\mathrm{obs}, \mathrm{peak}} \simeq 2^{-\beta-3} (\theta_\mathrm{obs} / \theta_j - 1)^{-2\alpha} L_{0, t_{j}}
\end{equation}
(Granot, Panaitescu, Kumar \& Woosley 2002, hereafter GP2002), where $\theta_j$ is the opening angle and $t_j$ is the jet break time for an on-axis observer. For $\beta = 0.5$ and $\alpha = 2$, which are the typical values of these parameters, we have that $L_{\theta_\mathrm{obs},\mathrm{peak}}  = 5.56\times10^{-2} L_{0, t_{j}}$. The peak time will be at

\begin{equation}
 T_\mathrm{peak} = [ 5 + 2\ln(\theta_{\mathrm{obs}}/\theta_j - 1)] (\theta_\mathrm{obs}/\theta_j - 1)^2 t_{j}~\mathrm{s}
\end{equation}
for the values above, we have $T_\mathrm{peak} \simeq 25 \times t_{j}$. Since $T_\mathrm{peak}  \simeq 90$~ks, $t_{j} \simeq 3.6$~ks.

Now, defining $a \equiv (1+\Gamma^2 \theta^2)^{-1}$, we have (GP2002)

\begin{equation}\label{a}
\nu(\theta_\mathrm{obs}) = a \nu (\theta=0) \\
;\\
F_{\nu} (\nu,\theta_\mathrm{obs},t) = a^3 F_{\nu} (\nu/a,0,at)
\end{equation}
where $\Gamma$ is the Lorentz factor.  At the peak time we have $\Gamma^{-1} \sim \theta_\mathrm{obs} - \theta_{j} = 2 \theta_j$.
By assuming $\theta$ is $\theta_\mathrm{obs} - \theta_{j}$, as GP2002 suggest, we have $a=0.5$ in the equations above.

The peak frequency for $\theta_\mathrm{obs} = 0$ is given by \\

\begin{equation}
\nu_\mathrm{M} = 3.3\times10^{14} (z+1)^{1/2} \epsilon_{B,-2}^{1/2}  \left(\frac{p-2}{p-1}\right)^{2} \epsilon_e ^2 E_{K,52}^{1/2} t_\mathrm{d} ^{-3/2} \mathrm{Hz} \\
\label{nu_m}
\end{equation}
where $t_\mathrm{d}$ indicates time in days. The maximum flux is \\

\begin{equation}
F_{\nu} (\nu_\mathrm{M}) = 1600 (z+1) D_{28}^{-2} \epsilon_{B,-2} ^{1/2} E_{K,52} n^{1/2} (t/t_{j})^{-3/4}~\mathrm{\mu Jy} \\
\label{flux}
\end{equation}
(Yost et al. 2003). $E_{K,52}$ is the kinetic energy of the ejecta, while $\epsilon_e$  and $\epsilon_{B,-2}$ are the fractions of shockwave energy given to radiating electrons and magnetic field respectively. $D_{28}$ is the luminosity distance of the burst, while $p$ is the index of the power law energy distribution of radiating electrons, $n$ the density in particles $\mathrm{cm}^{-3}$ of the circumburst medium. Subindices indicate normalized quantities, $Q_{x} = Q / 10^{x}$ in cgs units. Substituting the known parameters, taking $p=2.02$ to explain the flat X-ray spectrum, and remembering that for $\theta_\mathrm{obs} = 3\theta_j$ the observed $\nu_\mathrm{M}$ will be 1/2 of the $\nu_\mathrm{M}$ on-axis (see Eq.~\ref{a}), we have

\begin{equation}\label{power law}
F_{\nu} (\nu_\mathrm{i},\theta_\mathrm{obs},t_\mathrm{peak}) = 0.17 E_{K,52} ^{1.27} \epsilon_{B,-2}^{0.77} \epsilon_{e} ^{1.02} n^{1/2}~\mathrm{\mu Jy}
\end{equation}
where $\nu_\mathrm{i}$ is the flux in the $i'$ band ($3.9\times10^{14}$~Hz). At the peak of the rebrightening, we have $F_{\nu} \simeq200$ $\mu$Jy. Thus, we have the condition

\begin{equation}
E_{K,52}^{1.27} \epsilon_{B,-2} ^{0.77}  \epsilon_{e} ^{1.02} n^{1/2} \simeq 1200
\label{n}
\end{equation}

\subsubsection{Wide jet}\label{Wide}
An off-axis model cannot explain the early shallow decay if the observer has $\theta_\mathrm{obs} < \theta_j$; the observer must be slightly outside the opening angle of the outflow (i.e., $\theta_\mathrm{obs}$ a bit larger than $\theta_j$). The time when the afterglow emission begins its typical power law decay, $t\simeq 860$~s, can be taken as the epoch when $\Gamma^{-1} \sim \theta_\mathrm{obs} - \theta_j$. The following decay, with $\alpha \simeq 0.6$, can be explained if $\theta_\mathrm{obs} \simeq 3/2 \theta_{j}$ (Van Eerten et al. 2010). Finally, a steeper decay will be visible when the observer will see the radiation from the far edge of the jet, when $\Gamma^{-1} \sim \theta_\mathrm{obs} - \theta_j + 2 \theta_j = 5/2~\theta_j$. Assuming that $\Gamma \propto t^{-3/8}$, this second break would be seen at $t_{2} \simeq 5^{8/3} \times 0.86 \simeq63$~ks. However, at this epoch the afterglow is dominated by the narrow jet emission. It is important though that $t_2$ occurs before the end of the rebrightening, otherwise this model would predict a return to shallow decay once the rebrightening were over.
From Van Eerten et al. (2010), the brightness of an afterglow seen at 1.5~$\theta_j$ is $\sim 1/10$ of the brightness it would have if seen on-axis, in a given band. At 4500s, the $R$-band flux is $\simeq 100$ $\mu$Jy. If we assume $p=2.02$, we have 

\begin{equation}
E_{K,52} ^{1.27} \epsilon_e ^{1.02}  \epsilon_{B,-2} ^{0.77} n^{1/2} \simeq 13.3\\
\label{w}
\end{equation}
If we assume typical values $\epsilon_B=0.1$, $\epsilon_e=1/3$ and $n=10$ for both the narrow and wide jet, we obtain that the isotropic energetics of the narrow and the wide jet are $6.5\times10^{53}$ and $1.9\times10^{52}$ erg respectively. As for the half-opening angles of the outflow, a jet break at $\approx3.6$ks for the narrow jet would imply (Sari, Piran \& Halpern 1999) $\theta\simeq0.027$~rad. The opening angle of the wide jet is 2/3 as much as the observing angle, while the opening angle of the narrow jet is 1/3 as much; thus the wide jet opening angle will be twice that of the narrow jet. The beaming-corrected energies are $2.3\times10^{50}$ and $2.7\times10^{49}$~erg respectively. These values of the parameters are not unusual for GRB modeling.

In our model, the observed prompt $\gamma$-ray emission is dominated by the wide jet, since its edge is closer to the observer. To compute the prompt energy release in $\gamma$-rays that we would measure if we were within the opening angle of the wide jet, we can still use Eq. \ref{a}. However, we must consider that, during the prompt emission, $\Gamma$ is much higher than during the afterglow emission; opacity arguments (M\'esz\'aros 2006) and measurements (Oates et al. 2009) indicate that initially $\Gamma \gtsim 100$. Assuming $\Gamma = 100$, one obtains $a\simeq0.12$. Granot et al. (2005), in their note 6, suggest that for $\Gamma^{-1} < (\theta_\mathrm{obs} - \theta_\mathrm{j} ) < \theta_\mathrm{j}$, the fluence roughly scales as $a^{2}$.
Thus, an observer within the opening angle of the wide jet would detect a fluence $0.12^{-2} \times 1.2 \times10^{-5} = 8.25 \times 10^{-4}$ erg cm$^{-2}$. The corresponding energy emitted in $\gamma$-rays would be $E_\mathrm{iso} \simeq 5 \times 10^{54}$~erg. These values would already be very high. We know from our previous modeling, which takes into account the off-axis position of the observer, that the kinetic energy of the wide jet is $1.9\times10^{52}$~ erg. Thus, the efficiency in converting the initial jet energy into $\gamma$-ray photons would be $\eta = E_\mathrm{iso} / (E_\mathrm{iso} + E_{K,52}) \simeq 99\%$. This inferred extreme efficiency is rather difficult to explain for all models of prompt emission, and it constitutes a problem for the off-axis model. We note, however, that the strong decrease of the observed fluence with off-axis angle may come from the assumption of a sharp-edge jet. For a structured jet with an energy and Lorentz factor profile, one may lessen the difficulty inferred above. Moreover, a lower efficiency would be derived if the kinetic energy of the outflow were higher than $1.9\times10^{52}$ erg; in turn a higher kinetic energy is possible assuming different values of the parameters $\epsilon_B$, $\epsilon_e$ and $n$.

\subsubsection{Chromatic behaviour}\label{chromatic}
This modeling, however, does not yet take into account the presence of a spectral break during the rebrightening, which seems to cross the optical band from higher to lower frequencies. Such crossing may also explain the chromatic behaviour of the optical afterglow at the rebrightening. Taking into account equations (\ref{a}) and (\ref{nu_m}), which give the value of $\nu_\mathrm{M}$ as observed on-axis and how its value is modified by observing the outflow off-axis, we find the condition

\begin{equation}
E_{K,52} ^{1/2} \epsilon_{B,-2} ^{1/2} \epsilon_e ^{2} \simeq 4.2\times10^3
\end{equation}
The high value for the right-hand is needed to have $\nu_\mathrm{M}$ in the optical range $\sim 10^5$~s after the trigger, even from a largely off-axis observer.\\
Eq (\ref{n}) has to be modified, because we are now assuming that at the rebrightening we are observing the peak flux $F_{\nu_\mathrm{M}}$. It becomes

\begin{equation}
E_{K,52} \epsilon_{B,-2}^{1/2} n^{1/2} \simeq 28
\label{n2}
\end{equation}

To satisfy these equations together, one would need the isotropic energy $E_{K,52} \sim 10^7$ and a value of density of $n \sim 10^{-14}$, both unphysical. As a further consequence of these extreme values for the energetics and densities, the Lorentz factor of the jets is also enormous. In fact, in order to be decelerated at $t_\mathrm{obs} \simeq 900$~s in such a thin medium, the initial Lorentz factor of the jet should be (Molinari et al. 2007) $\Gamma \sim 30000$. For these reasons, the model of the two-component jet seen sideways cannot be considered viable if, during the rebrightening, there is chromatic evolution due to the transit of $\nu_\mathrm{M}$.

\subsection{Reverse Shock and Forward Shock interplay}

We now examine the possibility that some of emission of GRB100814A afterglow may be produced by the RS. We suppose that a process of energy injection, due to late shells piling up on the leading ones, lasts the whole duration of observations, producing a long-lived RS \citep{sm00, zm01, uhm07}. In such circumstances, the RS emission can be visible in the optical band and, under the right conditions, in the X-ray band as well. We explore two variants of this scenario. In the first, the early optical emission is RS, while the rebrightening and the X-ray emission is due to FS. In the second version, the RS generates the early optical and all the X-ray radiation we observe, while the the rebrightening is due to FS emission.

\subsubsection{Early optical from Reverse Shock, X-ray and optical rebrightening from Forward Shock}\label{RS-FS1}


In this scenario, the break frequency determined by fitting the 50~ks SEDs is the synchrotron peak frequency $\nu_\mathrm{M,\mathrm{FS}}$ of the FS which is, initially, above the optical band. When $\nu_\mathrm{M,FS}$ approaches the optical band, the peak of the FS starts to dominate over the RS emission and produces the rebrightening and the chromatic behaviour we observe.  After $\sim70$~ks, both X-ray and optical emissions are of the same origin, the FS.\\
In the following, we shall be using the formulation of \citet{sm00} (hereafter SM00) to predict the temporal evolution of the flux due to FS and RS. We assume that the circumburst medium density $n$ decreases with radius as $n \propto r^{{\it -g}}$,where $r$ is the radius reached by the shocks, while the mass $M$ of the late ejecta which pile up with the trailing shells obeys $M(>\Gamma) \propto \Gamma^{-s}$, where $\Gamma$ is the Lorentz factor of these late shells. This parameter, $s$, defines the energy injection into the ejecta (see also Zhang et al. 2006), which keeps the shocks (both reverse and forward) refreshed. The energy of the blast wave increases with time as $E \propto t^{1-q}$, where $q$ is linked to the parameter $s$ (Zhang et al. 2006). We note that SM00 take the approximation of a constant density throughout the shell crossed by the RS and do not take into account the $PdV$ (where $P$ stands for pressure and $dV$ the element of volume) work produced by the hot gas (Uhm~2011). Changes in the density and mechanical work should be taken into consideration in a more realistic scenario; we do that using numerical simulations (see below). However, this formulation enables us to use relatively easy closure relations that link the spectral and decay slopes to the parameter $s$ of energy injection and the density profile $g$ of the surrounding medium. At 4500~s, we assume $\nu_\mathrm{M,RS} < \nu_\mathrm{O} < \nu_\mathrm{C,RS}$, (where $\nu_\mathrm{O}$ is the frequency of optical bands) since $\nu_\mathrm{O} > \nu_\mathrm{C,RS} > \nu_\mathrm{M,RS}$ would imply an implausible index $p$ for the energy distribution of the electrons that produce the RS emission, $p \approx 1$. We also assume that the X-ray band is above the cooling frequency of the FS emission, i.e. $\nu_\mathrm{C,FS}$. To have spectral indices consistent with those observed, we assume $p_\mathrm{FS}=2.02$ and $p_\mathrm{RS}=2.20$ for the Forward and the Reverse Shock respectively. These values of $p$  would lead to spectral indexes $\beta_\mathrm{RS} = 0.60$ and $\beta_\mathrm{FS} = 1.01$, which are within $3\sigma$ of the spectral parameters obtained when fitting the various SEDs.
We find that a uniform medium, $g=0$, cannot explain both the X-ray and early optical decay slopes. In fact, the amount of energy injection which would make the X-ray decay match the observed value produces too shallow an optical decay. Conversely, less energy injection, which would make the optical match the observation, would produce too steep an X-ray decay. Similarly, in the case of a wind-like circumburst medium with $g=2$, the amount of energy injection needed to model the observed optical decay would make the X-ray decay too slow. Instead, there exist solutions for ``intermediate" profile density, $g=1.15$. Other similar cases, halfway between constant and stellar wind profiles, have been found in modeling of GRBs \citep{sth08}. For $g=1.15$, energy injection characterized by $s=2.75$ (or $q\simeq0.6$), requires the decay indices of the RS and the FS emissions to be $\alpha_{RS} = 0.58$, and $\alpha_{\mathrm{FS}} = 0.58$.

We can also test whether this model predicts the correct rise and the decay slopes at the rebrightening (see Fig. \ref{all_LCs} and Table 2). For $g=1.15$ and $s=2.75$, $\nu_\mathrm{M,FS} \propto t^{-1.28}$ and $F_{\nu}(\nu_\mathrm{M,FS}) \propto t^{0.15}$ (see SM00). This implies that $F_{\nu}(\nu < \nu_\mathrm{M,FS})$ will rise as $\propto t^{+0.57}$ and decay as $t^{-0.51}$, in agreement with what is observed, except for a slightly shallower rise than observed.

As for the steep decay at $t>2\times10^5$~s, assuming a sideways spreading jet and the same energy injection, the decay slope would be $\alpha\approx1.3$. This is not consistent with the observed X-ray and optical and may be an issue of the scenario at hand.  We note that numerical simulations (e.g. Zhang \& MacFadyen 2009; Wygoda et al. 2011; van Eerten \& MacFadyen 2012) of jet breaks indicate that the ejecta undergo little sideways spreading, but the decay slope can be very steep because of jet edge effects. A degree of energy injection can moderate this fast decay and perhaps reproduce the observed behaviour, although this may be difficult to prove quantitatively.

To summarize, this model naturally explains the presence of a break frequency at the optical rebrightening, and the chromatic behaviour as a consequence of the interplay of RS and FS. A similar two-component scenario has already been used to model a few \sw~GRBs (e.g. J\'elinek et al. 2006) and pre-\sw~GRBs (see Kobayashi \& Zhang. 2003b). However, in previous cases the RS was supposed to vanish within a few hundreds seconds; in the case of GRB100814A the RS emission can be long-lived due to the continuous process of energy injection.\\
The model explains also why the rise and decay slopes in different filters are consistent. It explains also why the optical rebrightening has no X-ray counterpart and why the decay steepens first in X-rays and then in the optical band: the jet break takes longer to appear in the optical than in the X-ray band, because at 200~ks $\nu_\mathrm{M,\mathrm{FS}}$ is still close to the optical range, while $\nu_\mathrm{X} \gg \nu_\mathrm{M,FS}$. The decay slopes before the rebrightening and during the rebrightening itself are also roughly accounted for. In this scenario the X-ray and optical rebrightening are due to the same component. Thus, they should exhibit the same global temporal behaviour. If we extrapolate the peak time - peak frequency trend to X-ray frequencies, the peak time of the X-ray emission should have been observed several hundreds of seconds after the trigger (see Section 4.1). This agrees with observations, since the X-ray plateau appears to have started at that epoch.
Finally, such a long lived RS scenario would produce a bright radio emission; radio observations started a few days after the trigger and managed to detect a measurable radio flux (see Section \ref{radio}).

However, a more serious issue we have yet to consider is whether $\nu_\mathrm{M,FS}$ can be in the optical band as late as $\sim90$~ks.  We compute the value of $\nu_\mathrm{M}$ from 860~s, the earliest epoch when the emission of the FS shock is recorded. Since GRB100814A may be an intermediate case between constant density and stellar wind environment, we carry out our test using both equations (1) and (2) of Yost et al. (2003). We take $p_{\mathrm{FS}} \simeq 2.02$. Having derived the value of $\nu_\mathrm{M,FS}$ at 860~s, we follow its temporal evolution according to SM00 for $g=1.15$ and $s=2.75$. We find that $\nu_\mathrm{M,FS} \propto t^{-1.28}$. Thus, at 90~ks, we would have
\begin{equation}
\epsilon_{B,-2} ^{1/2} \epsilon_e^2 E_{K,52} ^{1/2} \simeq 760\;.
\label{model2a}
\end{equation}
in the case of constant density and

 \begin{equation}
\epsilon_{B,-2} ^{1/2} \epsilon_e^2 E_{K,52} ^{1/2} \simeq 470\;.
\label{model2b}
\end{equation}
for stellar wind.

Even assuming very large values for $\epsilon_{B,-2}$ and $\epsilon_e$, $33$ and $1/3$ respectively at equipartition, we would still need $E_K \sim 10^{58}$ erg for the case of a stellar wind. Such large energy is not predicted by any models of the GRB central engine.
 
\subsubsection{Early optical and X-ray emission from RS, rebrightening from FS}\label{RS-FS2}

A more plausible variant of the previous model, which also keeps all the advantages described above, predicts that all the emission in the X-ray band is also produced by the RS, with $\nu_\mathrm{X} > \nu_\mathrm{C,RS}$, while the FS produces the rebrightening. In this case, we can choose a large value for the parameter $p$ of the Forward Shock, and this greatly eases the energy requirements. We find that for $g=1.25$, $s=2.65$, $p_\mathrm{RS} = 2.02$, $p_{\mathrm{FS}} = 2.85$, the predicted temporal slopes are $\alpha_\mathrm{O}=0.57$ before the rebrightening, $\alpha_\mathrm{X} = 0.60$; the slope of the optical rise is $-0.52$, while the successive decay between $\sim50$ and $\sim200$~ks would be $\alpha_\mathrm{O}=1.11$. All these values are within $2.8\sigma$ of the observed values, except the rise, which is slightly steeper than predicted, and decay slope after the rebrightening, which is slower than predicted. However, the decay slope may be shallower because $\nu_\mathrm{M,FS}$ is still close to the optical band and the model is approximated, thus we can consider this solution satisfactory. The spectral slopes are accounted for, too.

Equation (\ref{model2b}) becomes

\begin{equation}
\epsilon_{B,-2} ^{1/2} \epsilon_e ^2 E_{K,52}  ^{1/2} \simeq 0.58
\label{model3a}
\end{equation}
We can derive, as we did for $\nu_\mathrm{M,FS}$, another condition. The maximum flux $F_{\nu} (\nu_{\mathrm{M},\mathrm{FS}})$ has to be equal to the peak flux reached at $\simeq 90$~ks, which is $\simeq 200$ $\mu$Jy. We find 

\begin{equation}
\epsilon_{B,-2} ^{1/2}E_{K,52}  ^{1/2} A_{*} \simeq 1.6 \times 10^{-3}\;.
\label{model3b}
\end{equation}
Where $A_{*}$ defines the normalization of the density profile, i.e. $n = A_{*} r^{-g}$ (see Chevalier \& Li 2000). These equations have to be solved jointly. Assuming the typical $\epsilon_e=1/3$, $\epsilon_{B,-2} ^{1/2} E_{K,52} ^{1/2} \simeq 5.3$. If we take $\epsilon_{B,-2}=33$ as well (these values of the $\epsilon$ parameters are reached at equipartition) then $E_{K,52}\simeq0.86$ at the onset of the external shock and energy injection. The medium is thin, with $A_{\ast} \simeq 3\times10^{-4}$. \\
Using the values of $E_K$ and circumburst density we can also estimate the RS microphysical parameters. At $50~ks$ the X-ray emission is still dominated by the RS and, from the best fit model, there is a break at $92.4^{+42.6} _{-39.9} \mathrm{eV} \simeq 2.2\times10^{16} \mathrm{Hz}$, which must be the synchrotron cooling frequency $\nu_{C,RS}$. For the chosen values of $s$ and $g$, it decays as $t^{-0.06}$.  Thus, we can compute it at $t_\mathrm{break}=860$~s as a function of the relevant parameters, multiply it by $(50/0.86)^{-0.06} \simeq 0.76$ and force the result to be equal to the break energy we find at 50 ks.

For the value of $\nu_\mathrm{C,RS}$ at $t_\mathrm{break}$, which we have taken as the deceleration time $t_\mathrm{dec}$ (see Sect. \ref{opticaldata}), we adopt the formulation of Kobayashi \& Zhang (2003a), their Eq. 9,

\begin{equation}
\nu_\mathrm{C,RS} = 2.12\times10^{11} \left( \frac{1+z}{2} \right)^{-3/2} \epsilon_{B,\mathrm{RS,-2}} ^{-3/2} E_{K,52} ^{1/2} A_{*} ^{-2} t_{\mathrm{dec}}^{1/2} \, \mathrm{Hz}
\label{nu_C,RS}
\end{equation}

For the above values of density and energy it is $\nu_\mathrm{C,RS} = 4.7\times10^{19} \epsilon_{B,\mathrm{RS,-2}}^{-3/2}$~Hz. Thus Eq.\ref{nu_C,RS} implies a very high value for $\epsilon_{B,\mathrm{RS,-2}}$; taking $\nu_\mathrm{C,RS} \simeq 2.2\times10^{16} \mathrm{Hz}$ would imply $\epsilon_{B,\mathrm{RS,-2}} \approx 100$. Such value is very large and would imply a very strong magnetization of the outflow, for which the RS emission may be suppressed. However, the error on the break energy is quite large, with a $3\sigma$ upper limit of $0.45~\mathrm{keV}$. We can thus assume that $\epsilon_{B,\mathrm{RS,-2}} \gtsim 47$. Such limit indicates that the ejecta carry a considerable magnetic field; we caution that, in such condition, our analytical formulation may not be the most correct way to predict the dynamics and the flux produced by the RS (Mimica, Giannios \& Aloy 2009). However, for the sake of simplicity, we will assume that the theoretical derivation we have used so far still applies. In the following, we will assume $\epsilon_{B,\mathrm{RS,-2}} = 60$.
This value of $\epsilon_{B,\mathrm{RS}}$ derived above enables us to explain the spectral break at 4.5~ks as $\nu_{C,RS}$ too.


This model predicts the correct values for the late, post-jet break decay slopes, if one assumes that the jet is spreading sideways: from Table 1 of Racusin et al. (2009), for $p_{\mathrm{FS}}=2.85$, $q=0.6$, $\nu_\mathrm{O} < \nu_\mathrm{C,FS}$, the flux decays as $\alpha=2.07$, consistent with observations. Numerical simulations indicate that jets have little side ways spreading (see above) and the steep decay can be explained in terms of edge effect. However, by coincidence this effect seems to predict slopes consistent with those of the spreading jet model. As for the X-ray light curve, it is reasonable to assume that the RS emission post-jet slopes are similar to that of the FS after a jet break. Pressure and speeds of the RS and FS shocks should not change across the contact discontinuity that divides the two at the jet break time, so the sideways expansions due to overpressure in both regions should be similar and lead to comparable behaviours in terms of dynamics and related emission. Thus, the late X-ray decay slope can be explained by the model we are discussing.

We can now determine $\epsilon_{e,\mathrm{RS}}$. The optical flux at 860~s is RS emission, and the flux density is $F_{\nu} \simeq 300 \mu Jy$. The optical emission is

\begin{equation}
F_{\nu} (\nu_\mathrm{O}) = F \left( \nu_\mathrm{peak,RS} \right) (\frac {\nu_\mathrm{O}}{\nu_\mathrm{peak,RS}})^{-\beta}\;,
\label{optical1}
\end{equation}

where $\nu_\mathrm{peak} = \mathrm{max}(\nu_\mathrm{M,RS},\nu_\mathrm{SA,RS})$ \footnote{We do not know, at this stage, whether the peak flux of the RS will be reached at the synchrotron peak frequency or at the synchrotron self-absorption frequency}. Now, we know that
\begin{equation}
\nu_\mathrm{M,RS} = \Gamma^{-2} \nu_\mathrm{M,FS} \left( \frac{\epsilon_\mathrm{e,RS}}{\epsilon_\mathrm{e,FS}}  \right)^{2} \left( \frac{\epsilon_\mathrm{B,RS}}{\epsilon_{B,\mathrm{FS}}} \right)^{1/2} R_\mathrm{p} ^2\;,
\label{optical}
\end{equation}
where $R_p$ = $g_\mathrm{RS}/ g_\mathrm{FS}$ with $g=(p-2)/(p-1)$. We first find $\Gamma$ at the deceleration time, $\Gamma_{\mathrm{dec}}$, using Eq. 2 of Molinari et al. (2007), $A_{\ast} = 3\times 10^{-4}$ and $E_K = 0.86\times10^{52}$ erg. In this calculation and in the following ones, we assume, as stated previously, that $t_{\mathrm{break}}$ is the deceleration time of the leading shell. We find that $\Gamma_\mathrm{{dec}} \simeq 125$, weakly depending on density and $E$. For the values of the RS parameters already defined, and even assuming a very high value for $\epsilon_{e,\mathrm{RS}}=0.4$, we have $\nu_\mathrm{M,RS} < \nu_\mathrm{SA,RS}$ at deceleration time. Thus, the peak flux of the RS will be reached at $\nu_\mathrm{SA,RS}$ and in Eq. \ref{optical1} $\nu_\mathrm{peak}$ is the self-absorption frequency. We know that

\begin{equation}
F_{\nu} (\nu_\mathrm{peak, RS}) = \Gamma F_{\nu} (\nu_\mathrm{M, FS} ) \left( \frac{\epsilon_\mathrm{B,RS}} {\epsilon_{B,\mathrm{FS}}} \right)^{1/2}\;,
\label{RSpeak}
\end{equation}
where $\Gamma$ is the Lorentz factor at any given time\footnote{This condition is valid at any given time, not only at deceleration as usually assumed. The component moving at $\Gamma$ is responsible for the energy injection and just decelerates at the moment.}. For the values already found, we have $F_\nu (\nu_{\mathrm{peak,RS}}) = 2.2\times10^4~\mu$Jy at the onset of the deceleration. From the observed optical flux using Eq. \ref{optical1} we find $\nu_\mathrm{SA,RS} \simeq 9.8 \times10^{10}$~Hz. Together with other parameters, from Eq.~9 of SM00 we also find $\epsilon_\mathrm{e,RS}$, which is the only remaining unknown. We find that $\epsilon_\mathrm{e,RS} \simeq 0.19$.

We note that the observed spectral index in the optical $\beta_\mathrm{O}$ is not constrained toward low values at a few ks. Using multi filter GROND data, Nardini et al. (2014) find a value of $\beta_\mathrm{O} \sim 0.2-0.3$, which seems to decrease with time between $\sim 1$ and $\sim 10$~ks. Such value and behaviour cannot be explained in the standard external shock model, unless one assumes that the RS emission is in the fast cooling regime, $\nu_\mathrm{C} < \nu_\mathrm{O} < \nu_\mathrm{M}$, in a wind environment, so that $\nu_\mathrm{C}$ is rising. Since the synchrotron spectrum, around $\nu_\mathrm{C}$, is thought to be very smooth, one expects to see $\beta_\mathrm{O}$ to change from $\approx0.5$ to $\approx0$ when $\nu_\mathrm{C}$ approaches the optical band from redder frequencies. This configuration is not attainable in our scenario, in which the early emission is from RS. To estimate $\nu_\mathrm{M,RS}$, we start from $\nu_\mathrm{M,FS}$, and then use Eq. \ref{optical}. We know already that $\Gamma_\mathrm{{dec}} \simeq 125$. Thus, we have $ \nu_\mathrm{M,RS} \sim 8.7 \times 10^{9} $ Hz at 860~s with the values of $\epsilon_\mathrm{B,-2,RS}=60$ and $\epsilon_\mathrm{e,RS} = 0.19$. According to SM00, with $g=1.25$ and $s=2.65$ it is $\nu_\mathrm{M,RS} \propto t^{-0.81}$. Thus, at 4500 s it is $ \nu_\mathrm{M,RS}\simeq 2.3\times10^9$~Hz.  Even for higher values of $\epsilon_e$ of the RS, typical of a magnetized outflow, implausibly high values of $E$ or a much higher value of $p_\mathrm{RS}$ (which is however constrained to be $p_\mathrm{RS}<2.04$ by the X-ray spectral index) would be required to move $\nu_\mathrm{M,RS}$ above the optical band at 4500 s.

In our scenario, a more reasonable hypothesis to explain the spectral evolution between 1 and 10~ks is that, as time goes by, the second component producing the rebrightening becomes more and more important. This component has a blue spectrum ($\beta < 0$) in this phase, thus the observed SED, which is a sum of the two components, gets gradually shallower with time and mimics the observed $\beta_O$.

\subsection{Modeling of the Radio Emission.}

 We shall now investigate the behaviour of the radio light curves in the context of this scenario. The radio flux is still rising after the putative jet break, peaking at $10^6$ s and decaying afterwards. The rise of the radio flux can be ascribed to a few possibilities: i) the same component responsible for the optical peak moves into the radio band. However, if the optical peak at $10^5$ s is caused by the transit of $\nu_\mathrm{M,FS}$, for the same peak frequency to cross the radio band a few $10^9$ Hz at $10^6$ s, would require that $\nu_\mathrm{M,FS}$ should evolve as $t^{-5}$. This is not possible even in the context of a jet break.  ii) the radio peak marks the transit of $\nu_\mathrm{M,RS}$. 
At deceleration it is $\nu_\mathrm{M,RS} \simeq 8.7\times10^{9}$~Hz and decays as $t^{-0.8}$ for the chosen values of $s$ and $g$; at the jet break time $\nu_\mathrm{M,RS} \simeq 1.5\times10^{8}$ Hz, and it is likely to decay faster from this point. Thus, $\nu_\mathrm{M,RS}$ is not expected to transit in the $4.7$ and $7.9$~GHz bands as late as $10^6$~s.

We are therefore left only with the possibility that the radio peak is due to the self-absorption frequency $\nu_\mathrm{SA}$, either of the RS or the FS, crossing the radio band from bluer frequencies. According to the analytical solution of a sideways spreading jet, the flux below $\nu_\mathrm{SA}$ is expected to become constant after the jet break; however numerical simulations \citep{van11a} have shown that the flux can still increase if the observing frequency is $\nu<\nu_\mathrm{SA}$.

We will attempt to find an order of magnitude value of this parameter, since it is not easy to find its analytical expression for $0<g<2$. By adopting the $g=2$ case of Yost et al. (2003) and considering a very tenuous medium (see above), the self-absorption frequency of the FS is expected to be at $\sim2\times10^5$~Hz at $1.3\times10^5$~s. After the jet break, it is not expected to rise within this time up to $\sim10^9$~Hz, even in the case of energy injection. A similar result is derived if we use SM00, their Eq.~9, to obtain the value of $\nu_\mathrm{SA,FS}$ at the deceleration time of 860~s, and then we constrain its temporal evolution plugging a density profile\footnote{Derived from Eq.~2 of Sari \& M\'esz\'aros (2000)} of $n\propto t^{-0.64}$ and $E_{K,52}\propto t^{0.4}$. If $\nu_\mathrm{SA,FS}$ basically did not depend on $E_{K,52}$ for $g=1.25$ and we thus neglected this dependence, the self-absorption frequency would be even lower and make its transit in the radio band even more difficult to attain.
Instead, the self-absorption frequency of the RS could be in the right range. We know already that $\nu_\mathrm{SA,RS} \simeq 9.8 \times 10^{10}$~Hz at deceleration time. From this epoch, we compute its evolution assuming, as above, that $n\propto t^{-0.64}$ and $E_{K,52}\propto t^{0.4}$. Thus, $\nu_\mathrm{SA,RS} \simeq 3\times10^{9}$~Hz at jet break time. To estimate $\nu_\mathrm{SA,RS}$ from this epoch onwards, we assume that $\nu_\mathrm{SA,RS} \sim \Gamma^{8/5} \nu_\mathrm{SA,FS}$ (SM00). 
In the jet break regime without energy injection, $\Gamma \propto t^{-1/2}$, while $\nu_\mathrm{SA,FS}\propto t^{-1/5}$, thus $\nu_\mathrm{SA,RS} \propto t^{-1}$. Thus, at $10^6$~s, $\nu_\mathrm{SA,RS}$ should be $\simeq 0.4$~GHz. However, because of the ongoing energy injection, $\Gamma$ will decrease more slowly, and it is not unreasonable to assume that $\nu_\mathrm{SA,RS}$ is still in the GHz range. A similar result can be obtained from Eq.~2 of Yost et al. (2003), if we determine $\nu_\mathrm{SA,FS}$ at deceleration, follow its temporal evolution as above, and derive $\nu_\mathrm{SA,RS}$ by multiplying by $\Gamma^{8/5}$.  The peak flux, too, should be in the right range. For the values of $s$ and $g$ chosen, RS peak flux evolves as $t^{-0.16}$ until the jet break. After that, we use the relation $F_{\nu} (\nu_\mathrm{peak, RS}) \propto \Gamma F_{\nu} (\nu_\mathrm{peak, FS})$. In jet break regime, $\Gamma \propto t^{-1/2}$, while $F_{\nu} (\nu_\mathrm{peak, FS}) \propto t^{-1}$. The latter is proportional to $E_{K,52}^{1/2}$; since in our case $E_{K,52} \propto t^{0.4}$, it is reasonable to assume $F_{\nu} (\nu_\mathrm{peak, FS}) \propto t^{-0.8}$. Combining the two, we get  $F_{\nu} (\nu_\mathrm{peak, RS}) \propto t^{-1.3}$. At the radio peak time $10^6$~s, the RS peak flux is thus expected to be $\sim 700$~$\mu$Jy, similar to what derived from observations. We therefore conclude, from this qualitative discussion, that the radio peak may be produced by the transit of the RS self-absorption frequency in this band. The fact that the $7.9$~GHz light curve is initially much flatter than the $4.7$~GHz one (see Sect. \ref{radio}) might also be explained, as $\nu_\mathrm{SA,RS}$ is moving from bluer to redder frequencies. \\

\subsection{Comments on the physical parameters.}
There exists some degeneracy in the derived values of the physical parameters. Different pairings of $s$ and $g$ can account for similar decays in the X-ray, optical and radio afterglow bands. However, under the assumption that $\epsilon_{e,\mathrm{FS}} < 1/3$, we find $A_{\ast} < 3\times10^{-4}$\ from Eqn.s \ref{model3a} and \ref{model3b}. Values of $E_{K,52}$ much higher than $\simeq1$ would imply higher $\Gamma$ and $F_{\nu} (\nu_\mathrm{peak})$; $\nu_\mathrm{SA,RS}$ should have to be lower to explain the flux at deceleration. This could be obtained by increasing the value of $\epsilon_{e,\mathrm{RS}}$.\\
A value of $E_{K,52} \simeq 0.86$ may imply a rather high efficiency of the mechanism converting kinetic energy into the initial burst of $\gamma$-rays, $\eta = E_\mathrm{iso} / (E_\mathrm{iso} +E_{K,52}) \simeq 0.9$. Such value can hardly be obtained in most of the prompt emission models. However, it is worth noting that the value of $E_{K,52}$ is calculated at deceleration, when the energy injection begins. It is possible that the energy injection is due to trailing ejecta shells which have also produced the $\gamma-$ray emission. If this is the case, the efficiency should be calculated when the energy injection ends. In our model, this process goes on for at least until the last radio detection, $\simeq9\times10^6$~s; at this epoch, the kinetic energy associated to the blast wave will be $\simeq 3.4\times10^{53}$ erg. Thus the efficiency would be $\simeq 0.17$.  To compute the beaming angle $\theta_\mathrm{j}$ of the outflow, we use the condition $\Gamma^{-1} \simeq \theta_\mathrm{j}$ which holds at jet break time, $\simeq 1.33\times10^5$~s. At this epoch, $\Gamma \simeq 36$; thus $\theta_\mathrm{j} \simeq 0.028$. At the end of observations, the beaming-corrected value for the kinetic energy is $\simeq 1.4\times10^{50}$ erg, typical of other GRBs (Frail et al. 2001; Ghirlanda et al. 2007).

Another important feature of the scenario we are devising is the very low density of the environment, $A_{*} \simeq 3\times10^{-4}$, which corresponds to a mass loss rate of a few $\times10^{-9}$ solar masses year$^{-1}$ from the progenitor of GRB100814A. Comparably low values of $A_{*}$, however, are not unprecedented in GRB afterglow modeling (e.g. Cenko et al. 2011), and have been predicted for very low metallicity stars \citep{vin01}. For the value of $A_{*}$ at hand, the blast wave would reach densities comparable to the average density of the Universe at $z=1.44$ at $\sim 10^7$~s if it kept expanding radially. It is therefore possible that the density profile turns into a constant one before this happens, although the quality of late time data is not good enough to see the effects of this transition.

We now briefly discuss how our modeling changes if the actual deceleration time is earlier than 856~s (see Sect. 2.3). We tested the hypothesis that the actual deceleration time is half this value, i.e. $428$~s. We find that equations \ref{model3a} and \ref{model3b} would change slightly, and we would find slightly different values of $A$ and $E$ to satisfy both equations; other microphysical parameters relative to the FS would stay the same. However, an earlier deceleration time would imply an higher Lorentz Factor, $\Gamma_\mathrm{dec} \simeq 150$ rather than $\simeq125$ as in the previous case with $t_\mathrm{dec}\simeq856$~s; Eq. \ref{RSpeak} would thus imply a higher RS peak flux. The peak frequency for the RS would still be $\nu_\mathrm{SA,RS}$, but since it is inversely proportional to the deceleration time, it would be roughly twice the previous value. Taken together, these two differences would make an initial optical flux, at the deceleration time, too high and incompatible with observations. The only way to decrease $\nu_\mathrm{SA,RS}$ and thus the flux in the optical band would be to increase $\epsilon_{e,\mathrm{RS}}$, but it would have to be as high as $\epsilon_{e,\mathrm{RS}}\simeq0.5$, which is impossible because $\epsilon_{B,\mathrm{RS}}\simeq0.6$ already to make $\nu_\mathrm{C,RS}$ in the right range (see Eq. \ref{nu_C,RS}) and the sum $\epsilon_{B,\mathrm{RS}}+\epsilon_{e,\mathrm{RS}}$ cannot be more than 1. Acceptable solutions would be possible only if $t_\mathrm{dec} \gtsim 600$~s, and radio observations might be explained as well. We therefore conclude that, in our model, the deceleration time of the leading shell can occur before 856~s but not much before.\\

We summarize a description of different models proposed so far, with their advantages and problems, as well as values of the physical parameters, in Table \ref{tab_models}.

\subsection{Numerical simulations}\label{simulations}
\begin{figure*}
 \centering
 \includegraphics[width=0.75\textwidth, angle=-00]{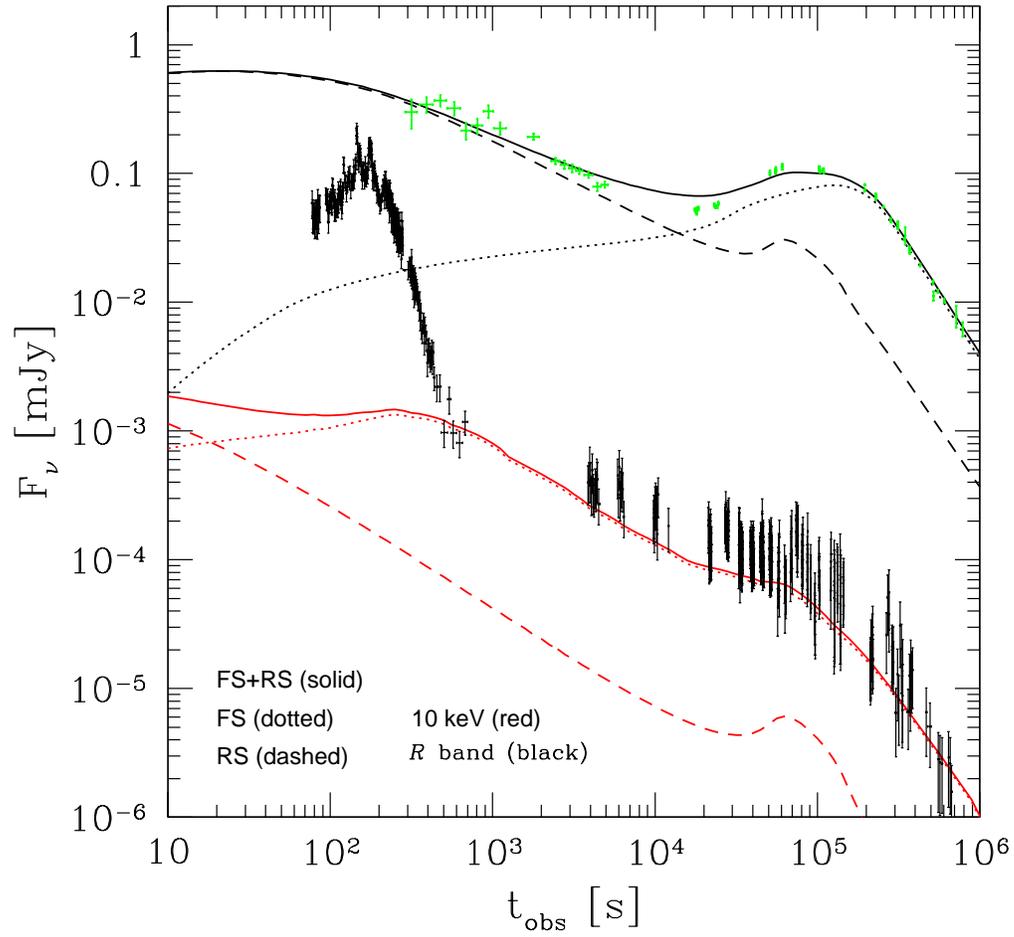}
 \caption{Results of the numerical simulation for GRB100814A X-ray and $R$ band light curves. RS stands for Reverse Shock, while FS stands for Forward Shock emission. The prompt emission is not described by this model. For more details see Sect. \ref{simulations}.}
\label{cRX}
\end{figure*}

\begin{figure*}
 \centering
 \includegraphics[width=0.75\textwidth, angle=-00]{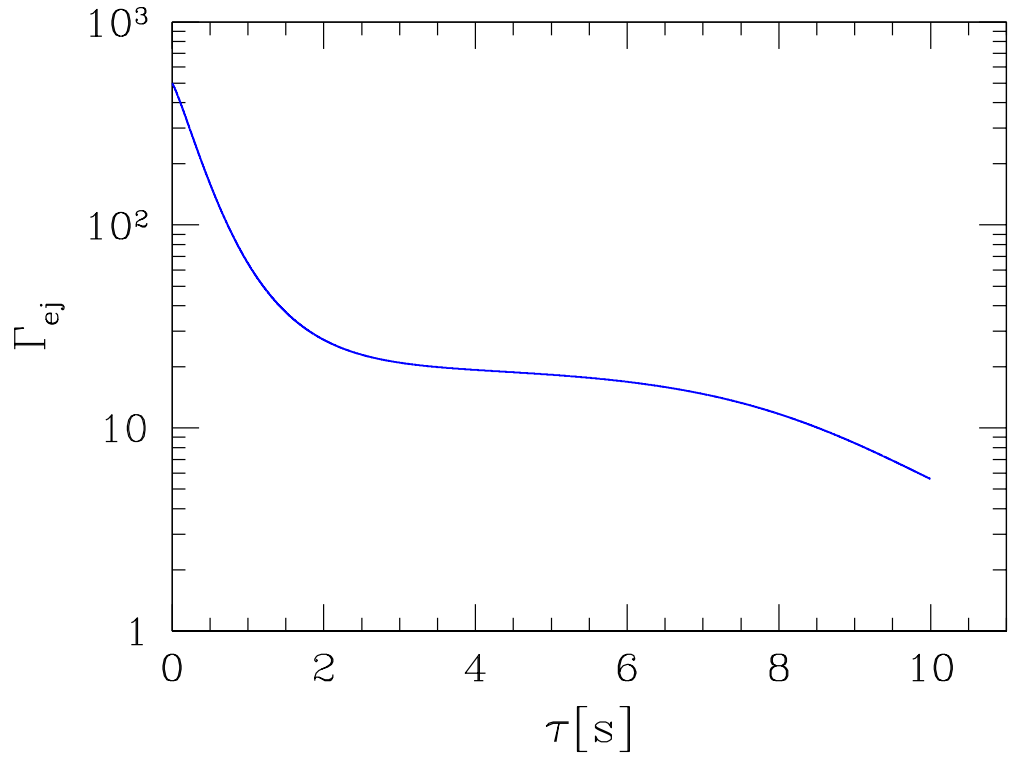}
 \caption{Distribution of Lorentz factor of the ejecta vs time of ejection $\tau$ in seconds. See Sect. \ref{simulations} for more details. }
\label{gej_tau}
\end{figure*}

We try now to approach the properties of GRB100814A using the numerical modeling of Uhm (2011) and Uhm et al. (2012). This is not based on full-blown hydrodynamical simulations, but a semi-analytical formulation of a relativistic blast wave. It applies the conservation laws of energy-momentum and mass in the region between the FS and the RS. Such work also considers a variable adiabatic index for the shocked gas in the regions intersected by the FS and the RS; this is quite important in the case of RS, which evolves from a non-relativistic regime to a mildly or relativistic regime as the blast wave propagates. We note that our simulations also make use of radial stratifications of the ejecta which can be quite different from a constant or a simple power law. Under such conditions, the FS dynamics may deviate from the self-similar solution of Blandford \& McKee (1976), but using the accurate numerical solutions of Uhm (2011), we can effectively predict the dynamics of the shocks.\\
 For the blast wave itself, we adopt a Lagrangian description (Uhm 2011, Uhm et al. 2012), which considers the blast wave as composed of many different Lagrangian shells all the way from the FS to the RS fronts. Each shell has its own physical parameters, such as energy density, radius, pressure, adiabatic index and, if necessary, magnetic field and electron energy distribution. This is rather different from the classical, simple analytical scenario of Sari et al. (1998), where the entire shocked region has the same radius, energy, pressure, magnetic field and power law distribution of electrons. The simulation shows the evolution of each shell, tracking the parameters such as energy, adiabatic index and magnetic field; it derives the minimum Lorentz factor and cooling Lorentz factor of the electrons  by solving the full differential equations (Uhm et al. 2012) numerically. Curvature effects of each shell, which has its own radius, are taken into account as well. Finally, the afterglow light curves are calculated by integrating the photons emitted from all shells that arrive at the same observer's time $t_\mathrm{obs}$.\\
 In the scenario we tested, the energy injection is due to late shells that collide with the trailing ones, and a long-lived RS develops.
The flux is due to both FS and RS, whose relative contribution evolves with time and depends on the observing frequency. For simplicity, we consider the flux in the $R$ and X-ray bands only and ignore the light curves in other optical and radio bands. The physical parameters involved were changed manually a few times, keeping some parameters fixed and altering others, until we found a visual good agreement between the derived light curve and the observations. We did not derive error margins. 
The results of the numerical modeling are shown in Fig. \ref{cRX}, and the distribution of the Lorentz factor of the ejecta versus time $\tau$ of the ejection is shown in Fig. \ref{gej_tau}. The FS has $\epsilon_{e,\mathrm{\mathrm{FS}}} = 0.1$, $\epsilon_{B,\mathrm{\mathrm{FS}}} = 0.01$; the RS has $\epsilon_\mathrm{e,RS} = 0.1$, while $\epsilon_\mathrm{B,RS} = 0.05$. Both shocks create a population of radiating electrons whose energy distribution is a power law with index $p=2.1$. The isotropic kinetic energy involved is $10^{54}$~erg, and the ambient medium density is 1 cm$^{-3}$. The $\sim2\times10^5$~s jet break is caused by a jet opening angle of 0.07 rad. To provide a satisfactory picture, the RS needs to energize 100\% of electrons of the ejecta while the FS is much less effective, providing energy only to $1.2\%$ of electrons of the medium it is moving into. The agreement between the predicted flux in the R and X-ray bands is subjectively good except for a slight ($\sim30\%$) overestimate of the optical flux at $\sim 25$~ks. The optical flux is due to declining RS emission up to $\sim10$~ks, when the FS emission takes over and dominates afterwards. The X-ray emission is always dominated by the FS, although a small increase in flux ($10\%$) in this band is visible around $80$~ks. This model also predicts a hardening of the spectrum around the peak time, as observed.\footnote{Relativistic hydrodynamic 1D and 2D simulations (Mimica et al. 2012) have shown under certain conditions the relativistic ejecta may undergo a total or partial {\it lateral collapse} and be (totally or partially) disrupted by the circumburst matter. 
If a fraction of the jet would be choked due to this effect, less energy may reach the working surface of the jet, leading to light curves different from those predicted without such jet disruption. 
Certainly, the possibility of the jet collapsing laterally depends on a delicate balance between the external medium ram pressure and the jet total pressure. In order to elucidate whether this effect is truly relevant or simply produces a small readjustment of the ejecta in the transversal direction one may require detailed 2D and 3D simulations, which are however out of the scope of this paper.}\\ 
 We emphasize how it is possible, on the basis of agreement between the synthetic results and observations, to constrain the temporal evolution of the Lorentz factor of the material emitted by the central engine. Such a method opens interesting opportunities to explain diverse behaviours in GRBs and understand better the physics of the central engine. The rebrightening of GRB100814A occurs at $\sim 1$ day; it shows a slow rise slope and it looks smooth. A few GRBs show a much faster rise. A possibility, envisaged in Uhm et al. (2012) and Uhm \& Belobedorov (2007), is that the central engine produces shells with a variety of Lorentz factors, evolving with time and more complicated than a simple power law. In these circumstances, it is possible to reproduce faster rises and decays which are otherwise difficult to explain with the external shock model.\\ We point out that our numerical simulations have confirmed the basic scenario drawn from the analytical model. In order to have the X-ray emission and the optical rebrightening produced by FS, with $\nu_\mathrm{M,FS}$ crossing the optical band as late as $\sim 1$~day and $p\simeq2.1$, one needs either an extreme value of kinetic energy imparted to the whole bulk of the emitting medium, or a more realistic value of kinetic energy somehow imparted only to a tiny fraction of the medium. It is not clear how one could attain either.

\vspace{-0.2cm}
\subsection{Other possibilities}

\setlength{\parindent}{10ex} We shall now briefly discuss other possible scenarios to explain the behaviour of GRB100814A, in connection with other GRBs showing the same phenomenology.

\subsubsection{Changes of the other microphysical parameters}
The fact that the rebrightening is not visible in the X-ray requires strong {\it ad hoc} assumptions regarding the evolution of these parameters, which makes the whole scenario contrived and implausible (Panaitescu et al. 2006; see, however, Filgas et al. 2011).

\subsubsection{End of energy injection.}
The rebrightening is produced when the energy injection, in form of late shells which pile up on the leading ones, ends, and bright FS and RS reverberate throughout the ejecta themselves (Zhang \& M\'esz\'aros 2002, Vlasis et al. 2011). Before and after the rebrightening, the emission comes only from FS of the leading shell.\\
\setlength{\parindent}{10ex}It has been found that the rebrightening is prominent, as in the case of 100814A, only if the ejecta are collimated. This would explain why we see, shortly after the rebrightening, a jet break and why the break times are not simultaneous. The spectral evolution observed during the rebrightening can be explained if we assume that a RS spectrum, with its peak frequency crossing the optical band, is outshining the FS emission. 
This model predicts a late radio peak, more or less simultaneous with the optical peak. However, we have no radio observations at the epoch of the optical peak, so this prediction could not be tested. The late radio peak, which  occurred $\simeq 13$ times later than the optical peak, was likely due to the behaviour of critical frequencies and dynamics.
\subsubsection{Internal dissipation emission}
We shall now discuss the possibility that the optical emission of GRB100814A is not being produced by external shocks, but it is an outcome of dissipation processes occurring inside the ejecta themselves.\\ 
 First, optical flares may have already been found in GRB afterglows (Roming et al. 2006, Swenson et al. 2013, Kopa\v{c} et al. 2013) and at least some of them are likely to be produced by internal dissipation processes, like their X-ray counterparts. Therefore, internal dissipation processes could generate late optical emission in GRBs. Second, in addition to GRB100814A, other events like GRB081029 (Nardini et al. 2011; Holland et al. 2012) and GRB100621A (Greiner et al. 2013) show sudden optical rebrightening towards the end of the X-ray slow decay phase. Another similarity to the case of GRB100814 is that the X-ray light curves of these GRBs do not seem to be altered much during the optical rebrightening: the flux in this higher energy band does not exhibit any clear analog rise. A difference is that, in these events, the rise of the optical flux is much steeper than in 100814A, approaching $\alpha_\mathrm{O} \simeq - 10$. Furthermore, there is spectral variability and, sometimes, rapid temporal variability during the rebrightening itself.\\ 
While a complicated distribution of Lorentz factor of the shells can reproduce slopes steeper than those detected for GRB100814A, it may be nevertheless difficult to explain such extreme slopes and variability in the context of external shock mechanism. Now, if what we see in GRB100814A is only a ``mild" version of the same phenomenon registered in other GRBs, one may thus need to abandon the external shock scenarios and study the behaviour in the context of internal dissipation models, in which fast variability is allowed by high bulk Lorentz Factors. The X-ray afterglow of GRB100814A is among the brightest of any observed by {\it Swift} during the end of the plateau phase (see Fig.\ref{ag_lc.eps}). According to Panaitescu \& Verstrand (2011), the X-ray afterglow of bursts with chromatic behaviour is on average brighter than that of bursts that do not show it. This might indicate that in these events the origin of at least the X-ray emission is not from the FS, but some other mechanism, such as internal dissipation.\\
A drawback of this scenario is that we do not yet understand well the behaviour of the internal dissipation emission. Thus, such identification is rather {\it ad hoc}, and not much susceptible to testing. The chromatic behaviour at the optical rebrightening of GRB100814A is not clearly accounted for, nor is the late steep decay similar to that observed in the X-rays.

\section{Conclusions}\label{conclusions}

We have reduced and examined an ample set of data on GRB100814A, observed by $\it{Swift}$, $\it{Fermi}$, and several ground optical and radio facilities. A prominent feature of this burst is an optical rebrightening, starting around $15 - 20$~ks after the burst trigger, which follows a typical early phase of slow decay of the flux. Such a rebrightening is not present in the X-ray light curve. However, when the optical rebrightening gives way to a steep decay, the X-ray light curve shows a break and a steepening as well. The radio emission, instead, peaks around $10^6$~s.\\
The optical rebrightening has a chromatic behaviour. This is already evident in the analysis of light curves; furthermore, a study of the spectral energy distributions shows a possible spectral break in the optical band, which is consistent with the transit of the synchrotron peak frequency $\nu_\mathrm{M}$ through it.\\
We have discussed a few models to interpret the behaviour of GRB100814A. The first model theorizes a double component jet; initially, both X-ray and optical emission are produced by a wide outflow component, seen just off-axis. A narrow component produces the optical rebrightening when its emission enters the line of sight of the observer. While this model can reproduce the temporal behaviour observed, the occurrence of a spectral break in the optical band at $\sim1$ day after the trigger would require an unphysical value of kinetic energy.\\
A second model assumes that the observed emission is a combination of a long-lived RS, caused by continuous energy injection in the form of late shells, and FS. For a configuration of the circumburst medium density profile and strength of energy injection, simple analytical calculations show that the X-ray emission and the optical rebrightening can be attributed to FS, while the RS produces the early optical shallow decay. The late steepening is due to a jet break. This model explains why the X-ray light curve shows no sign of the flux rebrightening seen in the optical, while it breaks to a steeper decay at an epoch similar to that of the optical. However, this model has again difficulty in explaining the presence of $\nu_\mathrm{M}$ crossing the optical band during the optical peak since it requires a very high value of energy $E$ of the ejecta. \\
More detailed, numerical calculations based on the the modeling of Uhm et al. (2012) indicate that the general behaviour can be described with the interplay of FS and RS, and more reasonable values of energy. Furthermore, this numerical modeling enables us to constrain how the Lorentz factor of the shells emitted by the GRB central engine evolves in time, thus shedding light on the still poorly known physics of this object. On the other hand, in the case at hand, one would require that the FS accelerates only $\simeq1\%$ of the electrons of the surrounding medium, which may be difficult to explain.\\
A variant of this model which keeps its advantages and sidesteps its problem is one in which all the emission, both in the X-ray and optical, is actually due to RS, while the optical bump is due to the emergence of a FS component with steep spectrum. In this case, a very high value of energy is not needed: $E\sim10^{50}$ erg after correction for beaming. 
Furthermore, this model predicts the correct optical post-jet break slopes if one assumes that the jet edge effect produces decay slopes similar to those expected for jets with sideways expansion.\\
The interplay between FS and RS emission may explain other GRBs that have an optical bump and chromatic behaviour. For different strengths of energy injection and density profile of the medium, a variety of behaviours, either chromatic or achromatic, can be reproduced. However, it is difficult to explain events which have a steep optical rebrightening with external shock scenarios. This is especially true when rapid flux fluctuations are present at the top of the rebrightening, for example in GRB081029 or GRB10621A. Therefore, a possibility we cannot exclude is that either or both the X-ray and optical emission are due to some internal dissipation mechanism.\\
GRB100814A belongs to the growing family of events whose afterglow cannot be explained by a simple component FS emission, but requires a superposition of more components, either produced by different regions of the ejecta or due to different blast waves. This category of events includes bursts with chromatic behaviour and rebrightenings at the end of the slow decline phase such as GRB100814A. Detailed temporal and spectral analyses of multi-wavelength data is needed in order to test the different scenarios, identify and characterize the different components present in afterglows. Thankfully, the combination of {\it Swift} and ground based facilities allows observers to produce an ample and extended coverage of GRBs and shed light on their complex and intriguing behaviour.

\section{Acknowledgements}

MDP, MJP,  NPK and SRO acknowledge United Kingdom Space Agency (UKSA) funding.
MDP thanks M. A. Aloy, F. Daigne and A. Mizuta for insightful discussions at "Supernovae and Gamma-Ray Burst 2013" conference, Kyoto.
CGM thanks the Royal Society, the Wolfson Foundation and the  Science and Technology Facilities Council (STFC) for support.
FG acknowledges support from STFC.
APB and PAE acknowledge UKSA support. This work made use of data supplied by the UK Swift Science Data Centre at the University of Leicester.
AG  acknowledges funding from the Slovenian Research Agency and from the Centre of Excellence for Space Sciences and Technologies SPACE-SI, an operation partly financed by the European Union, European Regional Development Fund and Republic of Slovenia".
MI, YJ, and S. Pak acknowledge the support from the Creative Initiative program, grant No. 2008-0060544 of the National Research Foundation of Korea (NRF), funded by the Korea government (MSIP).
SS acknowledges financial support from  support by a Grant of Excellence from the Icelandic and the Iniciativa Cientifica Milenio grant P10-064-F (Millennium Center for Supernova Science), with input from "Fondo de Innovaci\'{o}n para la
Competitividad, del Ministerio de Econom\'{\i}a, Fomento y Turismo de Chile", and Basal-CATA (PFB-06/2007).
The Liverpool Telescope is operated by Liverpool John Moores University at the Observatorio del Roque de los Muchachos of the Instituto de Astrofisica de Canarias. 
The Faulkes Telescopes, now owned by the Las Cumbres Observatory Global Telescope network, are operated with support from the Dill Faulkes Educational Trust.


\clearpage

\clearpage

\clearpage

\clearpage

\clearpage

\clearpage

\clearpage

\setcounter{table}{0}
\begin{table*}
 \caption{Overview of the observations.}
 \label{tab1}
 \begin{tabular}{@{}lllrl} 
\hline
{\bf Telescope or Observatory} & Telescope Aperture & Filter or freq./energy band & Notes \\
\hline 
EVLA   &        & 4.7, 7.9 GHz    & \\
BTA/Scorpio & 6.0 m   & $R,I$             &   \\
CQUEAN & 2.1 m   & $r$,$i$             &   \\
IAC80  & 0.82 m  & $R$               &   \\
CAHA   & 1.23 m  & $R$               &   \\
NOT    & 2.56 m  & $R,V,B$           &   \\
GTC    & 10.4 m  & $r$, R500B, R500R & 1 \\
LOT    & 1 m     & $g'$,$r'$,$i'$,$z'$     &    \\
LT     & 2.0 m   & $R$,$i$   & \\
FTN    & 2.0 m   & $R$,$i$   & \\
ROTSE  & 0.45 m  & unfilt & 2 \\
UVOT   & 0.30 m  &  $wh$,$v$,$b$,$u$,$uvw1$,$uvm2$,$uvw2$,gu &  3 \\
XRT    &        & 0.3-10 $\mathrm{keV}$ & \\
BAT    &        & 15-150 $\mathrm{keV}$ & \\
\hline
notes&&&\\
1	&- R500B, and R500R are spectroscopic observations \\
2	&- calibrated to $R$c \\
3	&- gu is uv grism \\ 
\hline

 \end{tabular}
\end{table*}

\clearpage

\begin{table*}
\centering
\begin{tabular}{llll}
\hline
\multicolumn{4}{l}{\textbf{$\mathbf 1^{\rm st}$ power law segment - chromatic evolution}}\\
\hline
Band	&	$\alpha_1$		&	$t_{\rm{b},\,1}$ (ks)	\\
\hline
$uvw2$	&$ 	\approx-0.60		$&$	\approx37.50			$\\
$uvm2$	&$	-0.66	\pm	0.29		$&$	51.39	\pm	7.92		$\\
$uvw1$	&$	-0.56	\pm	0.25		$&$	55.49	\pm	6.65		$\\
$u$		&$	-0.59	\pm	0.22		$&$	56.11	\pm	4.69		$\\
$b$		&$	-0.65	\pm	0.57		$&$	55.17	\pm	11.64	$\\
$v$		&$	-0.60	\pm	0.14		$&$	83.71	\pm	7.33		$\\
$g'$		&$	\approx-0.68		$&$	65.83	\pm	35.98	$\\
$white$	&$	\approx-0.69		$&$	\approx	48.66		$\\
$r'$		&$	-0.66	\pm	0.02		$&$	73.53	\pm	1.6		$\\
$R_c$	&$	\approx-0.68		$&$	74.92	\pm	1		$\\
$i'$		&$	-0.59	\pm	0.02		$&$	92.50	\pm	2.74		$\\
$z'$		&$	\approx-	0.68		$&$	\approx	79.39		$\\
\hline
\multicolumn{4}{l}{\textbf{$\mathbf 2^{\rm nd}$ and $\mathbf 3^{\rm rd}$ power law segment - achromatic evolution}}\\
Parameter& Value				& Parameter		& Value	\\
\hline
$\alpha_2			$&$	0.48\pm0.02	$&$	\alpha_3	$&$ 1.97\pm0.02	$\\
$t_{\rm{b},\, 2}\,(\rm{ks})	$&$ 217.7\pm2.4	$&$n_1=n_2		$&$ 10			$\\
\end{tabular}
\caption {Best fit parameters of the late-time optical afterglow. The fit yields $\chi^2_{\rm red} = 2.96$ with
126 d.o.f.. $n_1$ and $n_2$ are the smoothness parameters. Some light curve parameters could not be constrained for sparsely sampled light curves segments.  For those segments we only report fit estimates without the error.}
\label{tab:ag_fits}
\end{table*}

\vspace{-10cm}

\begin{table*}
\begin{center}
\begin{tabular}{cc}

Filter & $t_\mathrm{peak}$ (ks)\\
\hline
$w2$    &    $36.54 \pm 6.48$ \\
$m2$    &    $51.78  \pm 5.48$ \\
$w1$    &   $ 53.87  \pm  3.49$ \\
$u$       & $54.87   \pm 1.88$ \\
$b$       & $55.06  \pm  4.09$ \\
$v$       & $82.20  \pm  5.77$ \\
$white$  &  $48.96  \pm  6.33$ \\

$g'$  &   $  65.98  \pm  1.60$ \\
$r'$   &   $  75.33  \pm  1.54$ \\
$i'$    &  $  87.84  \pm  1.76$ \\
$z'$   &  $   80.22 \pm   3.79$ \\

\end{tabular}
\caption{Peak time of the rebrightening once the rise and the decay slopes have been fixed. The upper part of the table refer to UVOT light curves. See section \ref{opticaldata} for more detail.}
\label{tab_XX}
\end{center}
\end{table*}

\begin{table*}
\centering
\begin{tabular}{lrrrcrcrccc}
\multicolumn{1}{c}{Relation}	& \multicolumn{1}{c}{Slope}	& \multicolumn{1}{c}{normalisation}	& \multicolumn{2}{c}{Spearman's rank}		& \multicolumn{2}{c}{Pearson}				& \multicolumn{2}{c}{Kendall's $\tau$}	\\
											&											&													& \multicolumn{1}{c}{Value}	& Significance	& \multicolumn{1}{c}{Value}	& Significance	& \multicolumn{1}{c}{Value}	& Significance\\
\hline
$F_{\nu,\,\rm{p}} = N t^{-\alpha}$	&$0.87\pm0.40$	& $-5.09 \pm 1.98$		& $0.55\pm 0.25$	& $1.12\,\sigma$	& $0.55 \pm 0.25$	& $1.14\,\sigma$	& $0.44 \pm 0.23$	& $1.17\,\sigma$ \\
$F_{\nu,\,\rm{p}} = N \nu^{-\beta}$	&$-0.72\pm0.16$	& $9.93 \pm 2.39$		& $-0.81 \pm 0.11$	& $2.03\,\sigma$	& $-0.82 \pm 0.09$	& $2.15\,\sigma$	& $-0.70 \pm 0.14$	& $2.03\,\sigma$ \\
$\nu = N t^{-\gamma}$			&$-1.45\pm0.39$	& $21.85 \pm 1.93$		& $-0.80 \pm 0.12$	& $1.97\,\sigma$	& $-0.80 \pm 0.14$	& $1.93\,\sigma$	& $-0.65 \pm 0.16$	& $1.82\,\sigma$ \\
\end{tabular}
\caption{Correlation and linear regression analysis of the rising late-time optical afterglow. The linear regression was done in logarithmic space, i.e. $\log_{10} Y = \rm{N} + \rm{Slope} * \log_{10} X$.}
\label{tab:ag_correlations}
\end{table*}

\clearpage

\begin{table*}
\begin{center}
\begin{tabular}{cccccc}
\hline
                             & 500~s & 4.5~ks & 22~ks & 50~ks & 400~ks \\
\hline \\

Simple power law   	&           &             &            &            &               		      \\  
$E(B-V)$ (mag)         &            &             &           &           &  $(2.6\pm0.8)\times10^{-2}$   \\
$\beta  $                  	&           &             &            &            &   $0.95\pm0.01$   \\  
$\chi^2/$ d.o.f.     &           &            &             &            &   $58.7/45$           \\

\hline \\

Broken power law      		&            &             &              &            &               \\ 
$E(B-V)$  (mag) 		&   {\bf$(6_{-2.6} ^{+2.5})\times10^{-2}$} &    {\bf$(4.6^{+3.4} _{-2.8})\times10^{-2}$}   & {\bf$(0 ^{+0.2}) \times 10^{-2}$} & {\bf$(4.8^{+0.3} _{-1.8})\times10^{-2}$} & \\
$\beta_1$                  		&   $0.07^{+0.33} _{-0.30} $                     &  $ 0.52 ^{+0.07} _{-2.30} $                          &  $0.59^{+0.01} _{-0.19}$            &  $0.10^{+0.22} _{-0.04}$              &  \\ 
$E_{break} \mathrm{(eV)}$  &   $90.4 ^{+910} _{-86.8}$   			   &  $641 ^{+313} _{-640}$      					  &  $1240 ^{+230} _{-1170} $         &  $10.0^{+9.6} _{-3.8}$                  &  \\
$\beta_2$                             &   $0.89 ^{+0.13}  _{-0.05}$    			   &  $1.02 _{-0.08}$                 					  &  $0.93\pm0.09$                          &  $1.02 _{-0.02}$                           &  \\
$\chi^2/$ d.o.f.               &    $5.8/4$                            			   &  $11.8/14$                          					  &  $55.2/34$                                  &   $119.5/114$             				& \\

\hline \\

Broken power law          &           			&             &            &            &               \\ 
with $\Delta\beta =1/2$  &           		       &             &            &                &               \\ 
$E(B-V)$  (mag)     &   {\bf $(4.4 _{-1.4} ^{+2.5})\times10^{-2}$}   &   \bf{$(4.7^{+3.3} _{-2.8})\times10^{-2}$}  & {\bf $(0^{+1.5})\times10^{-2}$} & {\bf$(1.4\pm0.5)\times10^{-2}$} & \\
$\beta_1$               	     &   $0.34^{+0.06} $                                          &  $0.52  _{-0.06}$   & $0.39 ^{+0.13} _{-0.04}$                        				  &  $0.50^{+0.02} _{-0.04}$      & \\ 
$E_{break} \mathrm{(eV)}$ &   $583  ^{+515} _{-259}$ & $655  ^{+305} _{-390}$ & $86 ^{+193} _{-66}$          										  &   $47.1\pm22.0$  				& \\
$\beta_2$               	     &   $0.84^{+0.06}$               &  $1.02 _{-0.06}$                  &   $0.89 ^{+0.13} _{-0.04}$ 										  &   $1.00^{+0.02} _{-0.04}$     & \\
$\chi^2/$ d.o.f.                    &   $5.95/5$                         &   $11.8/15$                          &    $56.1/35$                       								  &   $123.4/115$                       & \\

\hline \\

Sum of two broken power laws          &           &             &            &            &               \\ 
$E(B-V)$  (mag)                       &            &             &           &    {\bf $(1.1^{+2.1} _{-0.4})\times10^{-2}$}    &  \\
$\beta_{1,I}$                            &           &             &            &  $-0.33$                            &               \\ 
$E_{break,I} \mathrm{(eV)}$  &           &             &            &   $4.1^{+0.5} _{-0.6}$      &               \\
$\beta_{2,I}$                            &           &             &            &  $8.5 ^{+unconstrained}_{-6.3}$  &    \\
$\beta_{1,II}$                           &           &             &            &   $0.52 _{-0.04}$                             &               \\ 
$E_{break,II} \mathrm{(eV)}$  &           &             &            &  $92.4^{+42.6} _{-39.9}$  &               \\
$\beta_{2,II}$                           &           &             &            &  $1.02 _{-0.04}$               &               \\
$\chi^2/$ d.o.f.                          &           &            &             &   $111.6/112$                    &               \\

\end{tabular}
\caption[]{Best fit values of the parameters of the models for the 500~s, 4.5~ks, 22~ks, 50~ks and 400~ks SEDs. The spectral index including the X-ray segment is forced between 0.84 and 1.02. In models for which $\beta_2 = \beta_1 +1/2$, $\beta_1$ is forced between 0.34 and 0.52 (see text for details).
Notes:\\
50~ks SED: the sum of two broken power laws model has the low energy spectral index fixed to -0.33 (in our convention; it is rising as $F\propto \nu^{1/3}$). 400~ks SED: the fits with broken power law models become indistinct from the simple power law one, since the break energy tends to 1~eV.}
\label{tab_SEDs}
\end{center}
\end{table*}

\begin{table*}
\begin{center}
\begin{tabular}{ccc}
\hline \\
Model & Advantages & Problems \\ 
\hline
Single component jet & Shallow decay and optical rise in principle explained     & Behaviour for energy injection and FS onset  \\
                                   & by several possibilities: energy injection, FS onset,       & cannot be chromatic in X and optical bands.   \\ 
                                   & transit of $\nu_\mathrm{M}$ through the optical band.    & Transit of $\nu_\mathrm{M}$ is not possible because the\\                             										     				                 &                                                                  & optical flux should evolve with a slope\\									
                                   &											               & between $t^{+0.5}$ and $t^{-0.25}$. The power law\\
                                   	&												        & index of radiating electrons is $p\sim2$, and one\\
						&										               & would require an unfeasibly high ejecta \\
						&											        & kinetic energy to have $\nu_\mathrm{M}$ in the optical  \\ 
						&                   							               & band as late as $\sim 1$ day. \\

\hline 
Single component jet with                      & Higher density may enhance the flux for $\nu<\nu_\mathrm{C}$,    & Simulations show that flux rebrightening \\
density rise in the circumburst medium & i.e. the optical band, and leave the flux for $\nu>\nu_\mathrm{C}$ & is not prominent if the blast-wave encounters \\
										    & unchanged; $\nu_\mathrm{C}$ may move into the optical band    & a density enhancement. \\
										    & and cause the chromatic behaviour.                                & \\
\hline
\hline
Two-component jet: wide outflow producing  & It has already been invoked and reasonable           & Chromatic behaviour during the  \\ 
the early optical and X-ray, optical                 & physical parameters are needed to explain             & rebrightening is not explained. \\
rebrightening and late X-ray from narrow      & the observed light curves. 						      &  Extremely high efficiency required.\\
jet observed off-axis        						&                                           & \\
& & \\
$p$ & 2.02 & \\
$\theta_\mathrm{j,wide}$, $\theta_\mathrm{j,narrow}$, $\theta_\mathrm{obs}$ & 0.054, 0.027, 0.081 \\
$\epsilon_{e}$, $\epsilon_{B}$, $n$ & 1/3, 0.1, 10 \\ 
$E_\mathrm{K,52,narrow}$, $E_\mathrm{K,52,wide}$ & $65$, $1.9$\\
$E_\mathrm{K,52,narrow,corr}$, $E_\mathrm{K,52,wide,corr}$ & $0.023$, $0.0027$\\
\hline
As above, two-component jet with $\nu_\mathrm{M}$  & Reasonable physical parameters are required  & Unreasonable kinetic energy of the narrow   \\			
transiting the optical band at 90~ks. & for the wide jet.								     & jet and circumburst medium density are  \\
 												    & 													&required. \\
$p$ & 2.02 \\
$\epsilon_{e}$, $\epsilon_B$, $n$ & 1/3, 0.1, $\sim10^{-14}$ \\ 
$E_\mathrm{K,52,narrow}$ & $\sim10^7$  \\
\hline
\hline
Interplay between RS and FS. Early optical & High value of $\nu_\mathrm{M,FS}$ explains why the      & With $p_\mathrm{FS} \leq 2.04$, an inconceivably high value\\
light curve from RS, all X-ray and optical     & rebrightening is present in the optical band but    & of kinetic energy of the ejecta is required to\\ 
rebrightening from FS.					   & not in the X-ray band. Same rise and decay         &  keep $\nu_\mathrm{M,FS}$ in the optical band $\sim1$ day after \\
									          & slopes in different filters during the rebrightening  & the trigger. Decay slope after the jet break is \\
									          & are accounted for. This model explains why the    & not correctly predicted.\\
									          & late X-ray and optical light curve show the similar & \\
									          & decay slopes and why the X-ray break is earlier.   & \\
                                                                         & Radio emission expected. 					    & \\
& & \\
      $s$             & 2.75   &\\
      $g$             & 1.15   &\\    
      $p_{RS}$    & 2.20  &\\
      $p_{FS}$    & 2.02  &\\
      $E_\mathrm{K,52}$ & $\sim10^6$  &\\
\hline
Interplay between RS and FS. Early optical & All the advantages above; the radio light curves & Optical rise slope during the  \\
light curve and all X-ray from RS, optical	    & and the late jet break slope are                            & rebrightening slightly under predicted. \\
rebrightening from FS.     					    & predicted too (with some   					       & \\
         &                                                           assumptions). & \\
      $s$          		& 2.65&\\
      $g$          		& 1.25&\\   
      $p_\mathrm{RS}$ 		& 2.02&\\
      $\epsilon_{e,\mathrm{RS}}$, $\epsilon_{B,\mathrm{RS}}$ & 0.60, 0.19&\\
      $p_\mathrm{FS}$ 		&  2.85 &\\
      $\epsilon_{e,\mathrm{FS}}$, $\epsilon_{B,\mathrm{FS}}$ & 1/3, 1/3 &\\             
      $E_\mathrm{K,52}$     		& 0.86 &\\
\hline        

\end{tabular}
\caption{Summary of models to describe GRB100814A: single jet (Sect. \ref{single}); two-component jet (Sect. \ref{Narrow}, \ref{Wide})
 two-component with chromatic behaviour (Sect. \ref{chromatic}); X-ray emission from FS, early optical from RS, rebrightening from FS (Sect. \ref{RS-FS1}); X-ray emission and early optical from RS, rebrightening from FS (Sect. \ref{RS-FS2}). Values of the parameters obtained in each section are also shown: $p$, index of the power law energy distribution of radiating electrons; $\theta_\mathrm{j,narrow}$ and $\theta_\mathrm{j,wide}$, opening angle of the narrow and wide components in the two-component jet model; $s$, energy injection parameter; $g$, slope of circumburst medium density profile; $\epsilon_e$ and $\epsilon_B$, fraction of the blast-wave energy going into radiating electrons and magnetic field respectively; $E_\mathrm{K}$, kinetic energy of the blast-wave; $E_\mathrm{K,corr}$, kinetic energy corrected for beaming.}
\label{tab_models}
\end{center}
\end{table*}

\clearpage


\clearpage


\begin{thebibliography}{}


\bibitem[\protect\citeauthoryear{Afanasiev \& Moiseev}{2005}]{scorpio} Afanasiev V.L., \& Moiseev A.V., 2005, PaZh, 31, 214; and 
Astronomy Letters, 31, 194. 

\bibitem[\protect\citeauthoryear{Akerlof et al.}{2003}]{Akerlof} Akerlof C.W. et al. , 2003, PASP, 115, 132.

\bibitem[\protect\citeauthoryear{Band et al.}{1993}]{ban93} Band D., Matteson J., Ford L. et al., 1993, \apj, 413, 281

\bibitem[\protect\citeauthoryear{Barthelmy et al.}{2005}]{bar05} Barthelmy S. D., Barbier L.M., Cumming J. R. et al., 2004, Space Science Reviews 120, 143

\bibitem[\protect\citeauthoryear{Blandford \& McKee 1976}{bl1976}]{blan76} Blandford. R. D., McKee C. F., 1976, PhFl, 19, 1130

\bibitem[\protect\citeauthoryear{Beardmore et al.}{2010}]{bea10} Beardmore A. et al., 2010, GCN 11087

\bibitem[\protect\citeauthoryear{Bertin}{2005}]{bertin} Bertin, E. SExtractor v2.5 Users Manual (draft), 2005, http://terapix.iap.fr/rubrique.php?id\_rubrique=91 

\bibitem[\protect\citeauthoryear{Beurmann et al.}{1999}]{beu99} Beuermann K. et al. 1999, A\&A 352, 26L

\bibitem[\protect\citeauthoryear{Bloom et al.}{2001}]{bl01} Bloom J. S., Frail D. A. \& Sari R., 2001, AJ, 121, 2879

\bibitem[\protect\citeauthoryear{Burrows et al.}{2005}]{ba05} Burrows D.N. et al., 2005, Space Science Review, 120, 165.

\bibitem[\protect\citeauthoryear{Cenko et al.}{2011}]{cen11} Cenko S. B., Frail D. A., Harrison F. A. et al., 2011, \apj, 732, 29

\bibitem[\protect\citeauthoryear{Chevealier \&Li}{2000}]{chl00} Chevalier R. A. \& Li, Z.-Y., 2000, \apj, 536, 195

\bibitem[\protect\citeauthoryear{Costa et al.}{1997}]{cos97} Costa E., Frontera F., Heise J. et al., 1997, Nature, 387, 783

\bibitem[\protect\citeauthoryear{De Pasquale et al.}{2006}]{dep06} De Pasquale M., Beardmore A., Barthelmy S. et al, 2006, MNRAS, 365, 1013

\bibitem[\protect\citeauthoryear{De Pasquale et al.}{2009}]{dep09} De Pasquale M., Evans P., Oates S. et al, 2009, MNRAS, 392, 153

\bibitem[\protect\citeauthoryear{Evans et al.}{2010}]{eva10} Evans P. A., Willingale, R., Osborne J. et al., 2010, MNRAS, 519, 102

\bibitem[\protect\citeauthoryear{Evans et al.}{2009}]{eva09} Evans P. A., Beardmore A., Page K. et al., 2009, MNRAS, 397, 1177

\bibitem[\protect\citeauthoryear{Evans et al.}{2007}]{eva07} Evans P. A., Beardmore A., Page K. et al., 2007, A\&A, 469, 379

\bibitem[\protect\citeauthoryear{Falcone et al.}{2006}]{fa06} Falcone A. D, Burrows D. N., Romano P., et al., 2006, AIP Conference Proceedings, 836, 386.

\bibitem[\protect\citeauthoryear{Filgas et al.}{2011}]{fil11} Filgas R., Greiner, J., Schady, P., et al., 2011, A\&A, 535 57

\bibitem[\protect\citeauthoryear{Frail et al.}{2001}]{fra01} Frail D. A., Kulkarni S. R., Sari R. et al., 2001, \apj, 611, 1005

\bibitem[\protect\citeauthoryear{Fukugita et al.}{1996}]{fukugita96} Fukugita M., Ichikawa, T., Gunn J.E., et al., 1996, \apj. 111, 1748

\bibitem[\protect\citeauthoryear{Galama et al.}{1998}]{gal98} Galama T. J., Wijers R. A. M., Bremer M. et al., 1998, \apjl, 500, 97

\bibitem[\protect\citeauthoryear{Gat et al.}{2013}]{gat13} Gat I., van Eerten H., \& MacFadyen A. 2013, \apj, 773, 2

\bibitem[\protect\citeauthoryear{Gehrels et al.}{2004}]{ge04} Gehrels N. et al., 2004, \apj, 611, 1005


\bibitem[\protect\citeauthoryear{Ghirlanda et al.}{2007}]{ghi07a} Ghirlanda G., Nava L., Ghisellini G. et al., 2007, A\&A, 466, 127

\bibitem[\protect\citeauthoryear{Ghisellini et al.}{2007}]{ghi07b} Ghisellini G., Ghirlanda G., Nava L. et al., 2007, \apj, 658L, 75


\bibitem[\protect\citeauthoryear{Golenetskii et al.}{2010}]{gol10} Golenetskii S. et al., 2010, GCN 11119

\bibitem[\protect\citeauthoryear{Granot et al.}{2005}]{gra05} Granot J., Ramirez-Ruiz E., Perna R., 2005, \apj,  630, 1003

\bibitem[\protect\citeauthoryear{Granot et al.}{2002}]{grb02} Granot J.,  Panaitescu A., Kumar P, Woosley S. E., 2002, \apj 570L, 61

\bibitem[\protect\citeauthoryear{Greiner et al.}{2013}]{gre13} Greiner J., Kruehler T., Nardini M. et al., 2013, A\&A, 560, 70

\bibitem[\protect\citeauthoryear{Guidorzi et al}{2014}]{gui14} Guidorzi C., Mundell C.G., Harrison R. et al. 2014, MNRAS, 438, 752


\bibitem[\protect\citeauthoryear{Holland et al.}{2012}]{hol12} Holland S. T., De Pasquale M., Mao J. et al., 2012, \apj, 745, 41

\bibitem[\protect\citeauthoryear{Huang et al.}{2005}]{hua05} Huang, K.Y., Urata, Y.,  Filippenko, A.V. et al.,  2005, \apjl, 628, L93 

\bibitem[\protect\citeauthoryear{J\'elinek et al.}{2006}]{jep06} J\'elinek M., Prouza M., Kub\'anek, et al., 2006,  \apj, 454L, 119

\bibitem[\protect\citeauthoryear{Jordi et al.}{2006}]{jordi06} Jordi K., Grebel E. K., Ammon K., 2006, A\&A, 460,339.

\bibitem[\protect\citeauthoryear{Kalberla et al.}{2005}]{kal05} Kalberla P. M. W., Burton W. B., Hartmann D. et al. 2005, A\&A, 440, 775

\bibitem[\protect\citeauthoryear{Kim et al.}{2011}]{kim11} Kim E., Park W.-K., Jeong H. et al. 2011, JKAS, 44, 115

\bibitem[\protect\citeauthoryear{Kinoshita et al.}{2005}]{LOT} Kinoshita D. et al., 2005,  ChJAA, 5, 315.

\bibitem[\protect\citeauthoryear{Kobayashi \& Zhang}{2003a}]{koz03} Kobayashi S. \& Zhang B., 2003a, \apj, 597, 455

\bibitem[\protect\citeauthoryear{Kobayashi \& Zhang}{2003b}]{koz03b} Kobayashi S. \& Zhang B., 2003b, \apjl, 582, 75

\bibitem[\protect\citeauthoryear{Kopac et al.}{2013}]{kop13} Kopa\v{c} D., Kobayashi S., Gomboc A. et al. 2013, \apj, 772, 73

\bibitem[\protect\citeauthoryear{Krimm et al.}{2010}]{Kri10} Krimm H.A. et al. 2010, GCN 11094

 \bibitem[\protect\citeauthoryear{Kuin et al.}{2015}]{kuin2015} Kuin, N.P.M., et al., 2015, ArXiv 1501.02433.

\bibitem[\protect\citeauthoryear{Liang et al.}{2007}]{Liang2007} Liang E.-W., Zhang B.-B., Zhang B. et al., 2007, \apj, 670, 565

\bibitem[\protect\citeauthoryear{Liang et al.}{2008}]{Liang2008a} Liang E.-W., Racusin J., Zhang B. et al., 2008, \apj, 675, 528

\bibitem[\protect\citeauthoryear{Liang et al.}{2013}]{lia12} Liang E.-W., Li L., Gao H. et al., 2013, \apj, 774, 13

\bibitem[\protect\citeauthoryear{M\'esz\'aros \& Rees}{1993}]{mer94} M\'esz\'aros P. \& Rees M., 1993  \apj, 418, 59.

\bibitem[\protect\citeauthoryear{M\'esz\'aros}{2006}]{mes06} M\'esz\'aros P., 2006  Rep. Prog. Phys., 69, 2259

\bibitem[\protect\citeauthoryear{Madau}{1995}]{mad95} Madau P., 1995, \apj, 441, 18

\bibitem[\protect\citeauthoryear{Mimica et al.}{2009}]{mim09} Mimica P., Giannios D., Aloy M. A., 2009, A\&A, 494, 879

\bibitem[\protect\citeauthoryear{Mimica et al.}{2012}]{mim12} Mimica P., Giannios D., Metzger B., 2012, EPJ Web of Conferences, 39, id.04003

\bibitem[\protect\citeauthoryear{Molinari et al.}{2007}]{Molinari2007a} Molinari E. et al., 2007 A\&A 469, L13

\bibitem[\protect\citeauthoryear{Nakar \& Granot}{2007}]{nag07} Nakar E. \& Granot J. 2007, MNRAS, 380, 1744

\bibitem[\protect\citeauthoryear{Nardini et al.}{2011}]{nar11} Nardini M., Greiner J., Krueler, T. et al., 2011, A\&A , 531, 39

\bibitem[\protect\citeauthoryear{Nardini et al.}{2014}]{nar14} Nardini M., Elliott, J., Filgas, R. et al., 2014, A\&A , 562, 29

\bibitem[\protect\citeauthoryear{Nishioka et al.}{2010}]{nis10} Nishioka Y. et al., 2010, GCN 11134

\bibitem[\protect\citeauthoryear{O'Meara et al.}{2010}]{ome10} O'Meara J. et al. 2010, GCN 11089

\bibitem[\protect\citeauthoryear{Oates et al.}{2009}]{oat09} Oates S., Page M.J., Schady P. et al., 2009, MNRAS, 395, 490. 

 \bibitem[\protect\citeauthoryear{Perley et al.}{2014}]{2014ApJ...781...37P} Perley, D.A., Cenko, S.B., Corsi, A. et al.,  2014, \apj, 781, 37

\bibitem[\protect\citeauthoryear{Park et al.}{2012}]{park12} Park W.-K. et al., PASP, 2012, 124, 839

\bibitem[\protect\citeauthoryear{Poole et al.}{2008}]{Poole} Poole T. et al., 2008, MNRAS, 383, 627. 

\bibitem[\protect\citeauthoryear{Panaitescu \& Verstrand}{2011}]{pav11} Panaitescu A. \& Verstrand T. 2011, MNRAS, 414, 3537

\bibitem[\protect\citeauthoryear{Panaitescu}{2008}]{pan08} Panaitescu A., 2008, MNRAS, 383, 1143

\bibitem[\protect\citeauthoryear{Panaitescu et al.}{2006}]{pan06} Panaitescu A., M\'esz\'aros P.,  Burrows, D. et al., 2006, MNRAS, 369, 2059

\bibitem[\protect\citeauthoryear{Press}{2002}]{Press2002a} Numerical recipes in C++ : the art of scientific computing, 2002, Press

\bibitem[\protect\citeauthoryear{Racusin et al.}{2009}]{rac09} Racusin J., Liang E.-W, Burrows D. et al., 2009, \apj, 698, 43

\bibitem[\protect\citeauthoryear{Rees \& M\'esz\'aros}{1994}]{rem94} Rees M \& M\'esz\'aros 1994  \apj, 693, 922.

\bibitem[\protect\citeauthoryear{Roming et al.}{2005}]{rom05} Roming P. et al., 2005, Space Science Reviews, 120, 95.

\bibitem[\protect\citeauthoryear{Roming et al.}{2006}]{rom06} Roming P., Van Den Berk D., Pal'shin, V. et al. 2006, \apj, 651, 985

\bibitem[\protect\citeauthoryear{Sakamoto et al.}{2009}]{sak09} Sakamoto T. et al., 2009 \apj, 693, 922.

\bibitem[\protect\citeauthoryear{Sari \& M\'esz\'aros}{2000}]{sm00} Sari R. \& M\'esz\'aros P., 2000, \apj, 535, 33L

\bibitem[\protect\citeauthoryear{Sari et al.}{1998}]{spn98} Sari R., Piran T. \& Narayan R., 1998, \apj, 497, L17

\bibitem[\protect\citeauthoryear{Sari et al.}{1999}]{sar99} Sari R., Piran T. \& Halpern J. 1999, \apj, 519L, L17

\bibitem[\protect\citeauthoryear{Sari \& Piran}{1999}]{sap99} Sari R. \& Piran T. 1999, \apj, 138 537

\bibitem[\protect\citeauthoryear{Schady et al.}{2007}]{sch07} Schady P., Mason K.O., Page M. et al., 2007, MNRAS, 377, 273

\bibitem[\protect\citeauthoryear{Schady et al.}{2010}]{sch10} Schady P., Page, M. J, Oates, S.R. et al., 2010, MNRAS, 401, 2773

\bibitem[\protect\citeauthoryear{Schlegel et al.}{1998}]{sch98} Schlegel D. J., Finkbeiner D. P. \& Davis M. 1998, \apj, 500, 525

\bibitem[\protect\citeauthoryear{Schulze et al.}{2011}]{Schulze2011a} Schulze S., Klose S., Bjornsson G. et al., 2011, A\&A, 526, 23

\bibitem[\protect\citeauthoryear{Smith et al.}{2002}]{JSmith} Smith J. A. et al., 2002, AJ, 123, 2121. 

\bibitem[\protect\citeauthoryear{Starling et al.}{2008}]{sth08} Starling R.L.C., van der Horst A., Rol E. et al., 2008, \apj, 672, 433 

\bibitem[\protect\citeauthoryear{Steele, et al.}{2004}]{LT} Steele I.A., Smith R.J., Rees P.C. et al., 2004, Ground-based Telescopes. Edited by Oschmann, J. M., Jr.,  Proceedings of the SPIE, 5489, 679.

\bibitem[\protect\citeauthoryear{Swenson et al.}{2013}]{swp13} Swenson C., Roming P., De Pasquale M. et al., 2013, \apj, 774, 2

\bibitem[\protect\citeauthoryear{Uhm \& Belobedorov}{2007}]{uhm07} Uhm Z.-L.\& Belobedorov 2007, ApJ, 665, L93

\bibitem[\protect\citeauthoryear{Uhm}{2011}]{uhm11} Uhm Z.-L. 2011, ApJ, 733, 86

\bibitem[\protect\citeauthoryear{Uhm & Zhang}{2012}]{uhm12} Uhm Z.-L., Zhang B. et al. 2012, \apj, 761, 147

\bibitem[Urata et al.(2014)]{2014ApJ...789..146U} Urata, Y., Huang, K.,  Takahashi, S., et al.\ 2014, \apj, 789, 146 

\bibitem[\protect\citeauthoryear{Van Eerten et al.}{2010}]{van10} van Eerten H., Zhang W., \& MacFadyen A., 2010, ApJ 722, 235

\bibitem[\protect\citeauthoryear{Van Eerten et al.}{2011}]{van11a} van Eerten H., Meliani Z.; Wijers R.A.M. J. et al. 2011, MNRAS, 410, 2016

\bibitem[\protect\citeauthoryear{Van Eerten et al.}{2011}]{van11} van Eerten H., \& MacFadyen et al. 2012, \apj, 751, 155

\bibitem[\protect\citeauthoryear{Varian}{2005}]{Varian2005a} Varian H. 2005, Mathematica Journal, 9, 768

\bibitem[\protect\citeauthoryear{Vlasis et al}{2011}]{vla1} Vlasis A., van Eerten H.,  Meliani Z. et al. 2011, MNRAS, 415, 279

\bibitem[\protect\citeauthoryear{Vink et al.}{2001}]{vin01} Vink J.S, De Koter A., \& Lamers H.J. 2001, A\&A, 369, 574

\bibitem[\protect\citeauthoryear{Von Kienlin et al.}{2010}]{vok10} von Kienlin A., on behalf of the {\it Fermi}-GBM team, 2010, GCN 11099

\bibitem[\protect\citeauthoryear{Yost et al.}{2003}]{yos03} Yost S., Harrison F. A., Sari R. et al., 2003, \apj, 597, 459

\bibitem[\protect\citeauthoryear{Wygoda et al.}{2011}]{wyg2011} Wygoda, N., Waxman, E. \& Frail, D.A., 2011, \apjl, 738, 23



\bibitem[\protect\citeauthoryear{Zhang \& M\'esz\'aros}{2001}]{zm01} Zhang B. \& M\'esz\'aros P., 2001, \apj, 552, 35L

\bibitem[\protect\citeauthoryear{Zhang \& M\'esz\'aros}{2002}]{zm02} Zhang B. \& M\'esz\'aros P., 2002, \apj, 566, 712

\bibitem[\protect\citeauthoryear{Zhang et al.}{2006}]{zha06} Zhang B., Fan Y.-Z.; Dyks J. et al., MNRAS, 2006, \apj, 642, 354

\bibitem[\protect\citeauthoryear{Zhang \&MacFadyen}{2009}]{zmf2009} Zhang W.-Q. \& MacFadyen, A. 2009, \apj, 698, 1261


\end{thebibliography}
\end{document}